\documentclass[10pt,journal,compsoc]{IEEEtran}
\usepackage{hyperref} 
\usepackage[caption=false,font=normalsize,labelfont=sf,textfont=sf]{subfig}
\usepackage{caption2}
\usepackage{algorithmic}
\usepackage{algorithm}
\usepackage{textcomp}
\usepackage{stfloats}
\usepackage{url}
\usepackage{verbatim}
\usepackage{graphicx}
\usepackage{cite}
\usepackage{bm}
\usepackage{amssymb}
\usepackage{amsmath,amsfonts}
\usepackage{array}
\usepackage{booktabs}

\ifCLASSINFOpdf
\else
\fi
\begin{document}
%
\title{Contactless Electrocardiogram Monitoring \\with Millimeter Wave Radar}
%
%
%
%


\author{Jinbo Chen, Dongheng Zhang, Zhi Wu, Fang Zhou, Qibin Sun, and Yan Chen
\IEEEcompsocitemizethanks{\IEEEcompsocthanksitem 
J. Chen, D. Zhang, Z. Wu, F. Zhou, Q. Sun, Y. Chen are with the school of Cyber Science and Technology, University of Science and Technology Of China, Hefei 230026, China (E-mail: jinbochen@mail.ustc.edu.cn, dongheng@ustc.edu.cn, wzwyyx@mail.ustc.edu.cn, fangzhou1958@mail.ustc.edu.cn, qibinsun@ustc.edu.cn, eecyan@ustc.edu.cn)\protect\\
}
}

%
%

\markboth{}%
{Shell \MakeLowercase{\textit{et al.}}: Bare Demo of IEEEtran.cls for Computer Society Journals}
%



\IEEEtitleabstractindextext{%

\begin{abstract}
The electrocardiogram (ECG) has always been an important biomedical test to diagnose cardiovascular diseases. Current approaches for ECG monitoring are based on body attached electrodes leading to uncomfortable user experience. Therefore, contactless ECG monitoring has drawn tremendous attention, which however remains unsolved. In fact, cardiac electrical-mechanical activities are coupling in a well-coordinated pattern. In this paper, we achieve contactless ECG monitoring by breaking the boundary between the cardiac mechanical and electrical activity. Specifically, we develop a millimeter-wave radar system to contactlessly measure cardiac mechanical activity and reconstruct ECG without any contact in. To measure the cardiac mechanical activity comprehensively, we propose a series of signal processing algorithms to extract 4D cardiac motions from radio frequency (RF) signals. Furthermore, we design a deep neural network to solve the cardiac related domain transformation problem and achieve end-to-end reconstruction mapping from RF input to the ECG output. The experimental results show that our contactless ECG measurements achieve timing accuracy of cardiac electrical events with median error below 14ms and morphology accuracy with median Pearson-Correlation of 90\% and median Root-Mean-Square-Error of 0.081mv compared to the groudtruth ECG. These results indicate that the system enables the potential of contactless, continuous and accurate ECG monitoring.
\end{abstract}

}

\maketitle

\IEEEdisplaynontitleabstractindextext

%
\IEEEpeerreviewmaketitle

\IEEEraisesectionheading{\section{Introduction}\label{sec:introduction}}

\vspace{-0.1cm}
\par The world has witnessed an increasing number of deaths caused by chronic and cardiovascular diseases (CVDs) \cite{bib1,bib2}. The Electrocardiogram (ECG) is one of the most significant biomedical signals, which is utilized to describe the activity of human heart and provides important information for the diagnosis of heart diseases. Experimental evidence has shown that many CVDs could be better diagnosed, controlled and prevented through continuous monitoring and analysis of ECG \cite{bib3,bib4,bib5}. Current approaches for ECG monitoring are based on body electrodes, which detect the small electrical changes that are a consequence of cardiac muscle depolarization followed by repolarization during each cardiac cycle. With advances in sensor and communication technology, continuous and real-time monitoring of ECG signals has been developed using various low-power biomedical sensors and energy efficient wireless transmission, which allows the lightweight system integration of miniaturized components attached to body directly \cite{bib6,bib7}. Although those electrode-based schemes could achieve continuous monitoring, practical experiences suggest that the nature of skin attachment in electric measurement restricts the reliability, adaptability and continuity of monitoring. Firstly, sticky electrodes and intrusive monitoring system might make patient feel uncomfortable, which decreases subjective willingness of wearing such device. Secondly, CVDs need long-time continuous monitoring of ECG to capture occasional cardiac disfunction event. However, the battery limitation and electrode falling deteriorate the continuity and reliability of the monitoring. Additionally, in some scenarios, such as for burned patients, highly infected patients, and premature babies, attaching adhesive ECG electrodes would not be feasible. To resolve the aforementioned challenges, wireless sensing innovation could play an important role in addressing contactless cardiac monitoring. However, early research in this area focused on monitoring heart rate \cite{zhang2020mtrack}, heart period \cite{bib9} or even more specific vibration signals from micro-cardiac events \cite{bib10} rather than direct ECG signals. In all, contactless ECG monitoring to identify patients at risk and provide accurate diagnoses is very important but currently still remains unaddressed.

\par Cardiac activity describes a series of events that take place in the heart. From the perspective of cardiac mechanical motion, cardiac activity represents repeating mechanical motions resulting in the sequence systole and diastole alternation over the atria and ventricle. This precise mechanical coordination ensures that blood is efficiently pumped by the heart and circulated throughout the body. Many techniques have been investigated to study this motion directly, including tagged magnetic resonance imaging (tagged MRI) \cite{bib11}, tissue Doppler imaging (TDI) \cite{bib12}, speckle tracking imaging \cite{bib13}, seismocardiography (SCG) \cite{bib14} and gyrocardiography (GCG) \cite{bib15}. From the electrical conduction point of view, cardiac activity represents the cycling transmission of electrical impulses from the sinoatrial node to the whole heart, which in turn results in the orderly repolarization and depolarization over atria and ventricle. This orderly pattern of electrical activity gives rise to the characteristic ECG measurement and tracing, which conveys a large amount of information about the structure and function of the heart.
\par Research focusing on cardiac mechanical and electrical activity have been investigated over years for cardiac monitoring purpose and plenty of studies have revealed that both methods provide essential information for clinical diagnosis and healthcare applications. Moreover, the research of the cardiac excitation-contraction coupling (ECC) mechanism \cite{bers2002cardiac,bers2001excitation} and its further modeling research \cite{gurev2011models,christoph2020inverse} not only provide a comprehensive understanding of the heart function in the cardiac electrical and mechanical activity integrated framework but also indicate there is a relation mapping between these two types of representation of cardiac activity.

\par In this study, we extend a new dimension of wireless sensing by designing a contactless ECG monitoring system with millimeter-wave radar that exploits the relation between cardiac mechanical activity and cardiac electrical activity. Specifically, the system depends on contactless sensing of cardiac mechanical activity and transforming it to the ECG representation. The radar is utilized to scan the heart by transmitting signals and receiving the reflections which are modulated by the cardiac motions. We design a series of signal processing algorithms to extract the 4D cardiac motion signals to comprehensively describe the cardiac mechanical activity. To do so, we firstly separate the reflections coming from different parts of torso by beamforming. Then we perform micro-motion amplification to eliminate the stronger motion interference (i.e. breath) and improve the SNR of cardiac motions. Next, the cardiac signal focusing and spatial filtering are performed to discover the cardiac reflections, suppress the random noise and finally extract the cardiac motion. To solve the ECG transformation problem, we propose a data-driven deep neural network with encoder-decoder architecture which leverages the temporal-spatial features of RF input and adapt the physiological mechanism of cardiac electrical conduction. 
\par We evaluate our system using commodity radar sensor and conduct 200 experimental trials over 35 participants, 4 different physiological status (normal-breath, irregular-breath, post-exercise and sleep). The total dataset consists 7200000 frames (10 hours) of radar measurements and its corresponding ECG ground truth. We evaluate the performance in terms of cardiac events timing accuracy and morphology accuracy. The timing analysis shows that our contactless ECG measurements can timing the Q-peaks, R-peaks, S-peaks, T-peaks with median error of 14ms, 3ms, 8ms, 10ms, respectively. The morphology analysis shows that our results achieve the median Pearson-Correlation of 90\% and median Root-Mean-Square-Error of 0.081mv compared to the ground truth ECG. The leave-one-out experiments indicate that the system performance over unseen patients achieves timing accuracy between 4ms-54ms,  3ms-9ms,  3ms-40ms 6ms-45ms  for  Q,  R,  S, T peak, RMS error 0.04mv-0.17mv, correlation between 65\%-97\%. In addition, the median error of 3ms, 90-percentile error of 9ms for timing R-R interval demonstrate the system potential for heart arrhythmias diagnosis.
\par The main contribution of this paper can be summarized as follows:
\par (1) To the best of our knowledge, we are the first to exploit the relation between cardiac mechanical activity and cardiac electrical activity through designing a contactless ECG monitoring system. The experimental results demonstrate that such a system enables the potential of contactless, long-term continuous and accurate ECG monitoring, which could facilitate the applications in a variety of clinical and daily-life environments.

\par (2) We propose to comprehensively describe the cardiac mechanical activity with 4D cardiac motion signals. To do so, we design a series of signal processing algorithms to extract the 4D cardiac motion signals from the raw RF signals.


\par (3) We propose to learn the domain transformation from the cardiac mechanical domain to the cardiac electrical domain. A hierarchical deep neural network is utilized to exploit the temporal and spatial features of RF input and incorporate the physiological domain knowledge. 

\par (4) We collect 10 hours of radar data and its corresponding ECG data from 35 participants under 4 different physiological status, which will be public to the research community. We believe that this dataset would facilitate future research of wireless sensing. 
\begin{figure*}[htbp]
\centering
\includegraphics[width=\linewidth,scale=1.00]{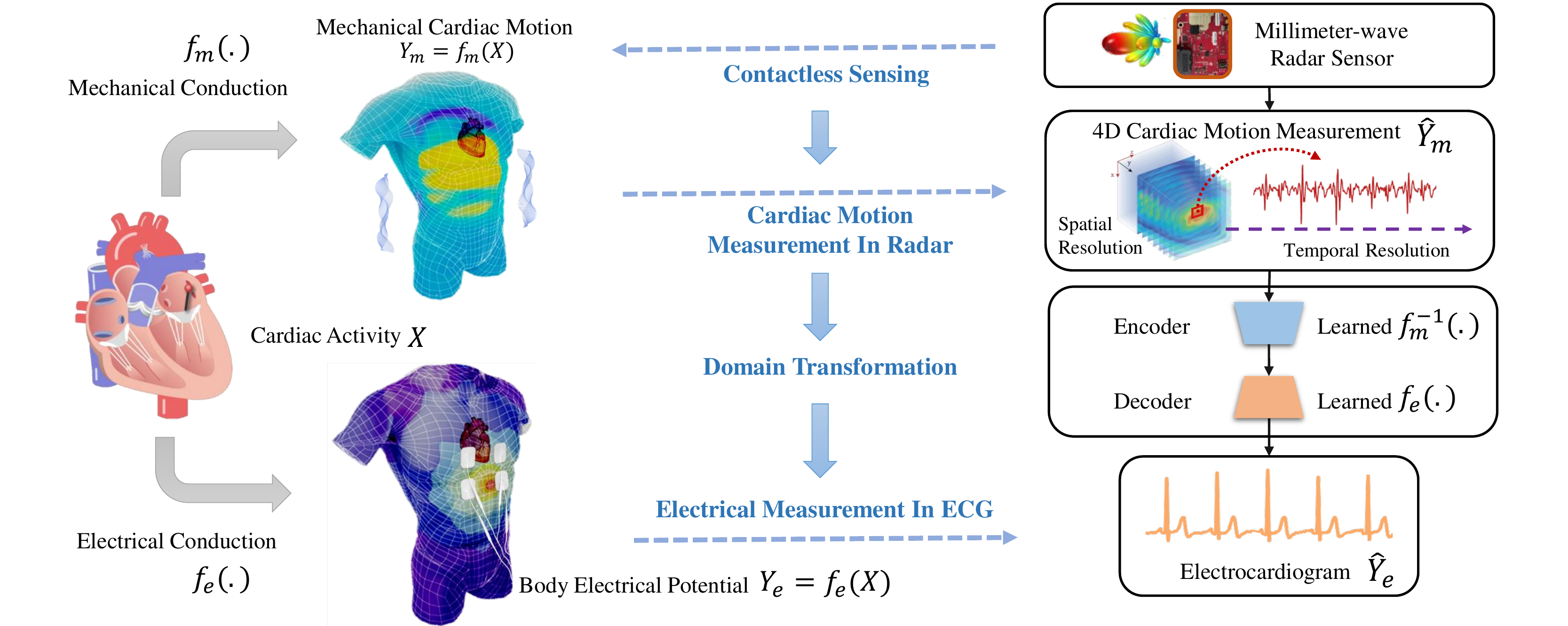}
\vspace{-0.5cm}
\caption{Working Principle and System Overview.}
\label{fig_sys}
\vspace{-0.5cm}
\end{figure*}

\vspace{-0.5cm}
\section{Related Work}
\par \textbf{ECG Monitoring:} Experimental evidence has shown that many of the CVDs could be better diagnosed, controlled, and prevented through continuous monitoring, as well as analysis of ECG signals \cite{bib3,bib4,bib5}. As a result, ECG monitoring systems have been developed and plenty of devices are being used in the healthcare sector for the past few decades.  Generally, these methods could be categorized into mobile-based, wearable-based, and sensor-based ones \cite{bib7}. Mobile-based devices for ECG monitoring involve a wide range of devices, including smartphones and pocket ECG monitors. \cite{aljuaid2020smartphone} evaluated the effectiveness of using a smartphone-based ECG monitoring device. Various research initiatives proposed to measure ECG with wearable devices. The authors in \cite{petrenas2015modified} proposed a modified Lewis ECG lead system for wireless monitoring of atrial arrhythmia, while \cite{fung2015electrocardiographic} explored a wearable on-body ECG patch to increase the possible device wearing time. Another category of ECG monitoring system utilizes innovative sensors to retrieve ECG signals without disturbing the patient’s comfort. \cite{rachim2016wearable} proposed a system consisting of capacitive coupled electrodes embedded in an armband. In \cite{ankhili2018washable}, washable long-term wearable sensors for fitness and activity monitoring is designed. The author of \cite{bouwstra2009smart} designed smart jacket-based continuous monitoring systems for prematurely born babies in the neonatal ICU.  Different from all aforementioned works which require sensor attached on the user, our proposed solution is based on millimeter wave radar, which does not require any contact between user and sensor and thus provides a more comfortable user experience. 

\par \textbf{RF-based Sensing:} Recent years have witnessed much interest in human sensing by RF signals. By leveraging the mechanism that body radio reflections are modulated by human motions, multiple RF sensing projects are dedicated for people localization \cite{chen2016time,chen2016achieving,han2016enabling,zhang2018multitarget}, monitoring walking speed \cite{chen2020speednet,hsu2017extracting}, gesture recognition \cite{li2021towards,zhang2021unsupervised}, events detection \cite{xu2017radio,xu2017trieds}, commodity WiFi sensing \cite{he2020wifi,zhang2019calibrating,chen2019residual}, human pose and skeleton estimation \cite{zhao2018rf, zhao2018through}. Previous works have also tried to monitor vital signs by RF sensing. \cite{adib2015smart} proposed an RF-based vital sign estimation system, which can accurately estimate the breath and heartbeat rate in the single-person scenarios. \cite{zhang2020mtrack} proposed MTrack system, which could measure the breath and heartbeat rate in the multi-person scenarios. \cite{zhao2016emotion} demonstrated extracting individual heartbeats from RF-signals in the emotion recognition task. Particularly, these works are limited to extracting the average heart rate or the heartbeat period. With the development of millimeter-wave radar, recent works focus on observing the cardiovascular events.  \cite{dong2020cardiogram} proposed a noncontact measurement of cardiogram with the diastole and systole periods of cardiac cycle. \cite{bib10} focused on recovering the morphology of the two dominant heart sounds (S1 and S2) found in PCG signals, achieving correlation coefficients of 80-82\% for each of them. More recently, \cite{ha2020contactless} contactless captured the SCG recordings and timed five key cardiac events with a median error between 0.26\%-1.29\%. \cite{wang2021mmhrv} measured the heart rate variability in daily-life environment with a median error of 28ms. The past works demonstrated the RF-based method could monitor the cardiovascular conditions with respect to the mechanical motion perspective. However, these works conducted solely from the perspective of cardiac mechanical activity (i.e. SCG, PPG), and thus were limited in large-scale clinical applications where ECG is still the golden standard to diagnosis CVDs. \par In this study, we extend a new dimension of wireless sensing to the ECG monitoring by exploiting the relation between cardiac mechanical activity and cardiac electrical activity. In contrast to the previous work, the 4D cardiac motion measurements in radar are proposed to represent the cardiac mechanical activity comprehensively. A deep neural network model is further proposed to solve the cardiac domain transformation problem to output the ECG with RF input.
\vspace{-0.25cm}
\section{System Design}
\subsection{Working Principle and System Overview}

\par Although the electrical signals are hard to be measured without the electrode attachment, the relation between the cardiac electrical and mechanical activities opens a window for ECG monitoring based on indirect measurements of cardiac mechanical activities. Specifically, this relation is driven by the cardiac ECC mechanism and their further body conduction process. In ECC, the ions within the heart bridge the electrical and mechanical representation of cardiac activity in a high nonlinear relation\cite{bers2002cardiac,bers2001excitation}. The experimental evidence has supported this correlation\cite{provost2011electromechanical,christoph2018electromechanical}. Also, the development of ECC modeling has demonstrated the forward/inverse mechano-electrical reconstruction can be resolved\cite{gurev2011models,christoph2020inverse}. From the further electrical conduction point of view, the cardiac electrical activities are conducted forward through the inner body and spread to the body surface, which could be measured in ECG\cite{gulrajani1998forward}. The forward problem can be modeled and calculated in a nonlinear projection by using biodomain model\cite{lines2003mathematical,boulakia2010mathematical}. From the mechanical conduction point of view, the pressure variation caused by heart mechanical motion pushes the damped elastic wave propagation to spread over the body surface resulting in the body surface mechanical motion, which also could be nonlinearly modeled depending on the viscoelastic properties in the thorax\cite{gurev2012mechanisms,laurin20153d}. 
\begin{figure*}[htbp]
\centering
\includegraphics[width=\linewidth,scale=1.00]{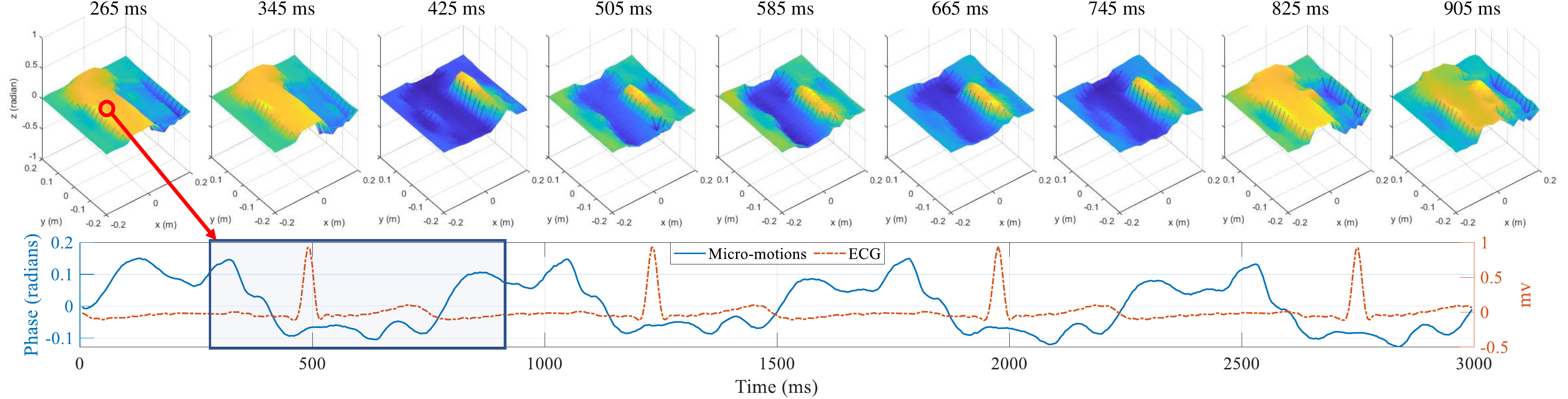}
\vspace{-0.5cm}
\caption{Radar measurements of body surface motion induced by the cardiac mechanical activity.}

\label{fig_concept_prove}
\vspace{-0.5cm}
\end{figure*}
\par In conclusion, as show in Fig.  \ref{fig_sys}, the ECC mechanism bridges the electrical and mechanical representation of homology cardiac activity $X$. Considering their further mechanical conduction and electrical conduction to the body surface, there is a nonlinear relation between the resulting electrical potential  $Y_{e}=f_{e}(X)$  and mechanical motion $Y_{m}=f_{m}(X)$ on the body surface. Inspired by the analysis above, our contactless ECG monitoring system is designed as illustrated in Fig. \ref{fig_sys}. Technically, our system mainly depends on contactless sensing of cardiac mechanical motion, modeling the cardiac mechanical-electrical conduction process and recovering the ECG information from the mechanical motion by domain transformation $Y_{e} = f_{e}(f^{-1}_{m}(Y_{m}))$.

\par Cardiac conducted mechanical motions are ranging with 0.2-0.5mm \cite{de1997chest}. It is also worth noticing that heart structure is often described as the size of 120mm in length, 80mm in width and 60mm in thickness approximately \cite{drake2005gray} and the mechanical motions induced from the cardiac cycle are distributed in various parts of the precordial area around chest \cite{lin2016identification}. Therefore, we not only need to measure the cardiac motions in precision of sub-millimeter but also should be capable of separating different motions spatially. 
\par To achieve contactless sensing, our solution is based on the radio frequency (RF) signals emitted by millimeter-wave radar. When the radio signals reach torso around chest, reflections occur, which modulate the cardiac motions to the echo signals. Specifically, we utilize a 77Ghz millimeter-wave radar with 4 Ghz bandwidth using a combination of a 2D antenna array and FMCW chirps to scan the entire 3D space for torso around the chest. The core idea to sense cardiac micro-motions with the radar sensor is to extract the phase of the received signal reflected from the target, which is given by the following equation:
\begin{eqnarray}
\phi(t) = 2\pi \frac{d(t)}{\lambda},
\end{eqnarray}
where $\lambda$ is the wavelength of the RF signals, $d(t)$ is the distance between the radar sensor and the position of reflection occurs, and $t$ is sensing time. The millimeter wavelength of 3.8mm endows high sensitivity to cardiac micro-motion. Also, this approach does not require participants to take off their clothes, since millimeter-wave signals can traverse through clothes. The 2D antenna arrary with horizontal and vertical components will be used by the beamforming in the next section to separate reflections coming from the different positions.

\par As a proof of concept, we demonstrate the radar measurement of surface motion that is induced by the cardiac mechanical activity in Fig. \ref{fig_concept_prove}. During the radar measurements, the subject is asked to hold their breath and remain quasi-static to avoid the interference from other motions like breath and body movements. The radar is placed above the precordial area within around 400mm. After 3D beamforming, the signals that have max reflected energy along the radar radial direction for different azimuth and elevation are selected to form the body surface reflection signals. The surface motions are extracted by the signal phase measurements which are linearly related to the radial distance variations between the torso surface and the radar sensor. In Fig. \ref{fig_concept_prove}, the body surface motions around throax within a cardiac cycle are represented in x-y plane for surface position and z axis for the corresponding phase measurement. It is clear that our radar sensor can observe the cardiac motions of the entire thorax with temporal-spatial resolution since the propagation of surface wave is observed during the atrial/ventricular systole and diastole states. When we look into the specific surface point marked by red circle, its corresponding phase variation over time is shown in Fig. \ref{fig_concept_prove}. We observe that the cardiac motion variation is consistent with the heartbeat cycle of ECG and the motion in different heartbeat cycles are similar with each other. From another perspective, the consistency of the periodicity of the cardiac motion for different positions of body surface and the consistency between the spatially distributed surface motion variation and the systole/diastole state in the cardiac cycle prove that our radar sensor can measure the cardiac motions with temporal-spatial resolution.

\par As for the domain transformation, technically, deep-learning method works based on multiple levels of representation, obtained by composing simple but non-linear modules that each transforms the representation at one level into a representation of more abstract feature level. With the composition of enough such transformations, very complex transformation functions can be learned. Therefore, we extend the boundary of deep-learning to the cardiac related domains transformation.

\par In general, we utilize the millimeter-wave radar sensor to measure the cardiac mechanical motions and the domain transformation is solved by the data-driven deep neural network. Once the training process finished, the network learned the mapping function and the radar cardiac motion measurements could be transformed to the cardiac electrical measurements in ECG. The key challenge with this idea is two fold: 
\par First, how to capture the cardiac motions with precision and robustness. The reflections of RF signals are very sensitive to the body posture. Even if a tiny body twist could make the reflections completely different. Also, the reflections of body are strongly subject dependent, which means there is no consistent characterization of body reflection signal from different subjects. Moreover, cardiac motions are within the sub-millimeter scale, making it difficult to be captured robustly in the presence of other sources of motion (like breath and body movement) in practice.

\par Second, how to design the network model to solve the transformation problem. The cardiac motion measurements in RF space are full of noise and distortion. Also the physiological domain knowledge of cardiac mechanical conduction and electrical conduction should be considered in the design of network model. 

\begin{figure*}[htbp]
\centering
\includegraphics[width=\linewidth,scale=1.00]{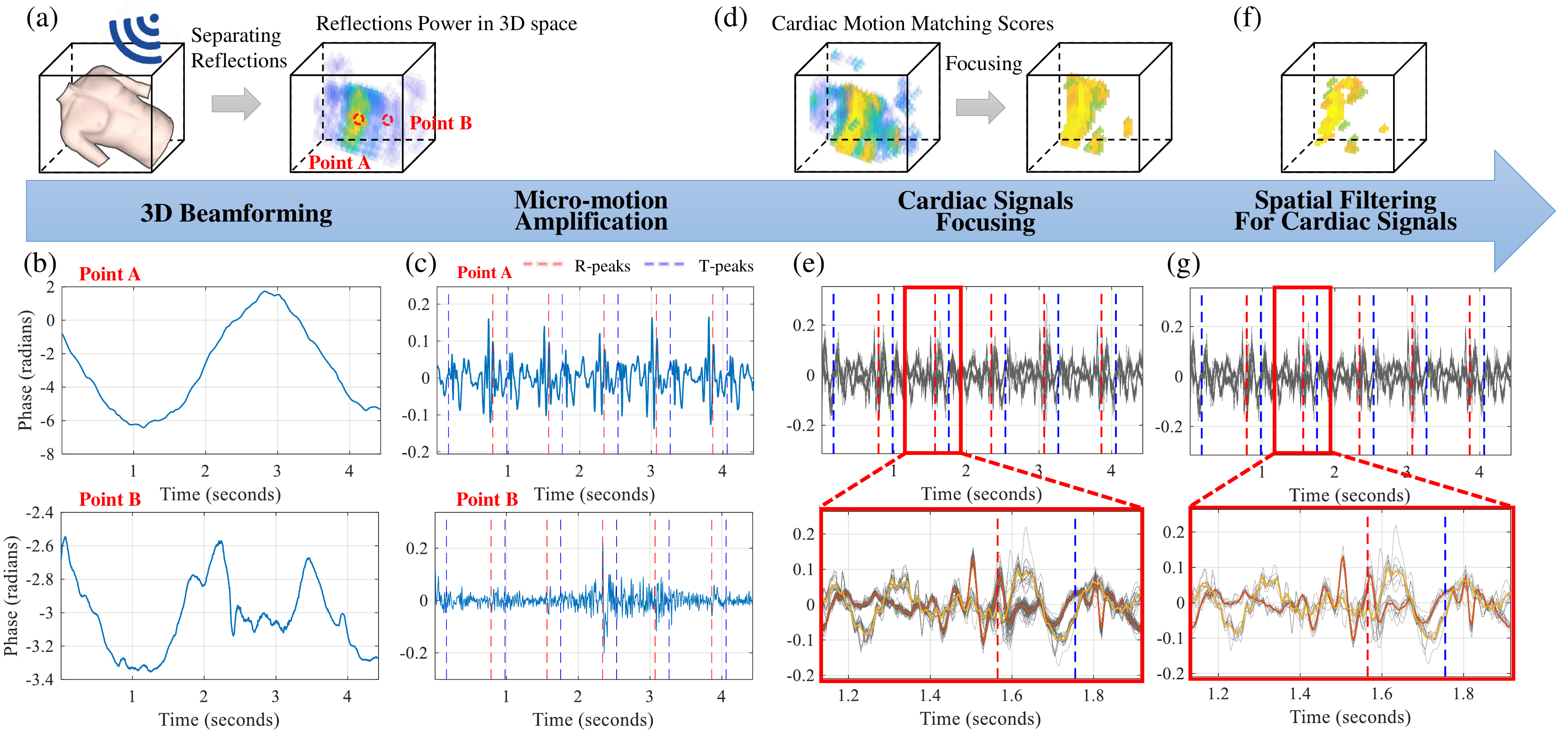}
\vspace{-0.2cm}
\caption{Signal processing flow for cardiac motion measurements: (a) The result of 3D beamforming; (b) Raw phase extraction from 2 different voxels where the distance to the heart varying from near to far away; (c) The result of micro-motion amplification for the voxel points A,B; (d) The result of pattern matching evaluation. Voxels color represents the matching score; (e) Morphology feature comparison of focused signals; (f) The spatial filtered signals distribution in 3D sensing space; (g) Morphology feature comparison of spatial filtered signals.}
\label{fig2-a}
\vspace{-0.4cm}
\end{figure*}

\par  To overcome these challenges, we present the design of our contactless ECG monitoring system with processing flow shown in Fig. \ref{fig_sys}, which incorporates two key modules: 
\par (1) \textbf{Cardiac Motion Measurements in Radar}: This module consists of a series of signal processing algorithms to robustly extract the cardiac motions with temporal-spatial resolution by separating the reflections from different positions of torso, amplifying the micro-motions, focusing the cardiac related reflections, and spatial filtering for cardiac signals respectively.
\par (2) \textbf{Deep Neural Network for Domain Transformation}: This module consists of a encoder-decoder based network architecture to solve the domain transformation problem. A hybrid Convolutional Neural Network (CNN)-Transformer based encoder is designed to exploit the temporal-spatial features of RF input. A temporal convolutional network (TCN) based decoder is designed to adapt the physiological mechanism of cardiac electrical conduction and model the conditional distribution of future ECG output.
\subsection{Cardiac Motion Measurements in Radar}

\par The goal of cardiac motion measurements in radar is to extract the motion signals reflected from the area where the cardiac mechanical activity happens. However, the reflections of human torso are very complicated, which are believed to be influenced by skin thickness and water content, resulting the quantitative variations in the torso reflectance among locations and individuals \cite{owda2020reflectance}. The presence of reflection variance might cause the interference between different reflected signals. This can be explained by considering that, the stronger reflection could leak the power to the surrounding weaker signals, modulate the motions and blur the information behind the signals, leading to the inaccuracy or even discontinuity of measurements. Moreover, the cardiac induced motion itself is masked by other motions like breathing, whose magnitude is significantly larger than the micro-motions induced from cardiac movements. Therefore, it is essential to measure the cardiac motions with reasonably spatial resolution and signal-to-noise-ratio (SNR). As shown in  Fig. \ref{fig2-a}, the proposed technique mainly consists of four signal processing algorithms: 3D beamforming, micro-motion amplification, cardiac signals focusing, and spatial filtering for cardiac signals, which will be explained in detail below respectively.
\vspace{-0.2cm}
\subsubsection{3D Beamforming}
\par Our radar sensor utilizes a combination of a 2D antenna array and FMCW to scan the 3D space for RF reflections. Each voxel in 3D space can be uniquely identified by its cartesian coordinates $(x,y,z)$. By projecting propagation path of the received signals into phase variations, the coherent combination of multiple phases provides a discriminative representation from a particular 3D voxel. Specifically, the signals received on all channels are combined using the following equation \cite{richards2014fundamentals}
\begin{eqnarray}
S(x,y,z,t) = \sum_{n=1}^{N}\sum_{t=1}^{T}y_{n,t}e^{j2\pi\frac{kr(x,y,z,n)}{c}t  }e^{j2\pi\frac{r(x,y,z,n)}{\lambda}  },    
\end{eqnarray}
where $N$ is the number of virtual channels formed by the pairs of transmitting and receiving antennas, $y_{n,t}$ is the channel signal received at time $t$, $k$ denotes the frequency modulation slope, $\lambda$ denotes the wave length and $r(x,y,z,n)$ denotes the round-trip distance through the voxel $(x,y,z)$ to the transmit-receive antenna pair in channel $n$.

\par After applying this 3D beamforming, we separate reflections coming from different parts of torso into different voxels, as shown in  Fig. \ref{fig2-a}(a). Also, each of projections is a time-series signal along the frame sequence. It then looks at the phase of the signal which is related to the traveled distance. In Fig. \ref{fig2-a}(b), we show different motion waveform sensed by radar distributed in different voxels.

\subsubsection{Micro-motions Amplification}
\par The variations in the phase correspond to the mixed motion caused by breathing, cardiac motion, and other random motion of human jointly. Note that the motions from torso around chest are predominant by the inflation and deflation of the lungs during the breathing cycle. These movements ranging from 4-12 mm with a frequency range of 0.2-0.34Hz (12-20 breaths per minute) \cite{de1997chest} are stronger and slower compared to the cardiac induced motions, which have amplitude of 0.2-0.5mm and higher frequency (heart rate is in the frequency range of 60-90 beats per minute) \cite{ramachandran1989three}. Therefore, the RF phase signals are mostly dominated by breathing. 
In such a case, one straightforward method to mitigate the influence of breath motion and extract the heartbeat signals is to utilize the frequency-domain filtering. However, the nature of the filter pre-setting manner, the signal distortion problem caused by mismatched filter settings, and nearby frequency power leakage, together lead to the difficulty in applying frequency-domain filtering method robustly. 
Note that the acceleration of breathing is smaller than that of heartbeats since breathing is usually slow while a heartbeat involves rapid contraction of the muscles. Thus, simliar to \cite{zhao2016emotion}, we can mitigate breathing and amplify the micro-motions by extracting acceleration signal as opposed to displacement directly. Considering the amplification of random system phase noise in high frequency band caused by the RF hardware during acceleration extraction, we utilize the differentiator with improved noise suppression based on least-squares smoothing in \cite{holoborodko2014noise} to extract the acceleration in lower frequency band while mitigate the high-frequency system noise:

\begin{eqnarray}
s_{0}^{\prime \prime} = \frac{(s_{-3} + s_{3}) + 2(s_{-2} + s_{2})- (s_{-1} + s_{1})-4s_{0} )}{16h^{2}},
\end{eqnarray}
where $s_{0}^{\prime \prime}$ refers to the second derivative at a particular time sample, $s_{i}$ refers to the value of the time series $i$ samples away, $h$ is the frame periodicity between consecutive samples. 

\par In Fig. \ref{fig2-a}(c), we demonstrate the micro-motion amplification process could improve SNR of cardiac motion, suppress the breathing effect and noise. When we look at the voxels (like point A) close to the human heart, breathing is dominant in RF phase signals and overwhelm the cardiac motion. In contrast, in the micro-motion amplified signal, clearly periodic pattern corresponding to each cardiac cycle could be observed. For signals (like point B) far away from the heart, the energy conducted by the heart's motion is consumed during the pathway, and the amplified signal has only meaningless noise.

\subsubsection{Cardaic Signals Focusing}
\par
\begin{figure}
\centering
\includegraphics[width=1\linewidth]{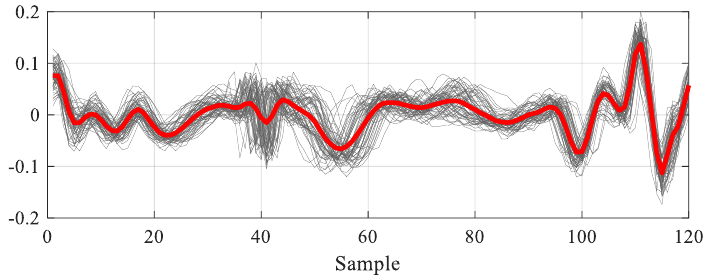}
\vspace{-0.4cm}
\caption{A voxel specific micro-motions are aligned by beats with ECG R-peaks timing. The aligned motions in grey are overlaid compared to the averaged trend in red color.}
\label{fig2-b}
\vspace{-0.4cm}
\end{figure}

 To capture the cardiac conducted motion completely, the 3D beamforming scanner is deployed in a redundant space for the torso. Therefore, we need to focus on the signals which carry the cardiac cycle information. The challenge is that we do not know exactly how the cardiac motions look like. To analyze the measured cardiac motion variation, we select a voxel which has similar periodic pattern of micro-motion compared with the cardiac cycle and aligned them into beats with the ECG R-peaks timing. The ensemble averaged trend and raw micro-motions over 80 seconds are overlaid and shown in Fig. \ref{fig2-b}. We observe that the morphology of averaged trend has a high similarity with the cardiac cycle aligned micro-motions. 
Therefore, we make an assumption that the morphology of cardiac motion is changing among subjects and locations, but the successive human cardiac cycle leads to similarity and periodicity of cardiac motion morphology from the same locations and same subjects. In other words, the cardiac motion related voxels should be those voxels whose time-domain micro-motion signals exhibit temporal similarity and periodicity. With such an intuition, we propose to utilize the periodicity based pattern matching method to identify the cardiac motion related voxels. 
\par Specifically, for time-series micro-motion signals, we identify the periodic morphology similarity by aligning the time series with a specific template, which can be stretched, to achieve a reasonable fit. Given the micro-motion measurements $S = s(1),...,s(n)$ with length $n$, we formulate the pattern matching process as solving the dynamic time warping (DTW) problem \cite{muller2007dynamic}. The pattern detection task involves searching a time series $S$ for instances of a template $T = t(1),...,t(m)$ with length $m$. A warping path $W = w_{1},...,w_{p}$ aligns the elements of $S$ and $T$, where each $w_{k}$ corresponds to a point $(i,j)_{k}$ indicating that $s_{i}$ is aligned with $t_{j}$. The DTW distance between $S$ and $T$ is defined as:
\begin{eqnarray}
{\rm DTW}(S,T) = \mathop{\min}_{W}\sum_{k=1}^{p}\delta(w_{k}),
\end{eqnarray}
where
\begin{equation}
\begin{aligned}
 \delta(w_{k})&=\delta(i,j)\\
 & = (s(i) - t(j))^2.  
\end{aligned}
\end{equation}
Let  $\overline{S} = \{\overline{S}_{1},\overline{S}_{2},...,\overline{S}_{l}\}$  denote the partition of sequence $S$ into non-overlapping contiguous segments. The final periodicity based cardiac motion pattern matching score can be stated as follows:

\begin{eqnarray}
{\rm P}(S) = \mathop{\min}_{T,\overline{S}}  \frac{1}{l}\sum_{\overline{S}_{i}\in \overline{S}} {\rm DTW}(T,\overline{S}_{i}).
\end{eqnarray}
However, the optimization problem above is hard to solve since the simultaneously optimization of $\overline{S}$ and $T$ leading to the exponential complexity. We simplify this iterative optimization into 2 sequential sub-problems solving: segments updating and template updating.
\par Segments updating is simplified by the human cardiac related prior. First, the valid segmentation length can be constrained within $[h_{min},h_{max}]$, which corresponds to the valid range of human heart beat duration. Considering the 200HZ radar sample rates and valid human rate in between 60 BPM to 120 BPM, $h_{min} = 100$ (120BPM/0.5 seconds) and $h_{max} = 200$ (60BPM/1 seconds). Second, given the periodic morphology-similar signals with the periodicity constrain $[h_{min},h_{max}]$, we can easily reach the prior that any segment with $h_{max}$ length of signal has at least one periodic pattern in itself. And this segment can be periodic aligned with the sequence and form an overlapping segmentation. Moreover, the start indexes of every two consecutive overlapping segments could be utilized to align new non-overlapping segmentation and represent the real periodicity of signals.
\par Inspired by the above prior information, we simplify segments updating optimization as follows: (1) Divide the original motion signals into multiple non-overlap $h_{max}$ length of segments as the coarse template candidates $T_{C}$; (2) Rewrite the optimization problem into overlapping segmenting version and calculate the best overlapping matching segmentation $\overline{S}^{*}$ based on the given coarse templates candidates $T_{C}$:
\begin{equation}
\begin{aligned}
&  \overline{S}^{*} = \mathop{\arg\min}_{\overline{S}} \sum_{\overline{S}_{i}\in \overline{S}} {\rm DTW}(T,\overline{S}_{i}),  \\
&{\rm s.t.} \ \ \ \ \ \   h_{min} \leq \tau(\overline{S}_{i+1}) - \tau(\overline{S}_{i})\leq h_{max} ,\\ 
& \ \ \ \ \ \ \ \  \ \  \ |\overline{S}_{i}|= h_{max},T \in T_{C},\\
\end{aligned}
\end{equation}
where $\tau(\overline{S}_{i})$ and $|\overline{S}_{i}|$ denote the start index and the length of segment $\overline{S}_{i}$ respectively. The constraint terms denote that the length of real heartbeat is valid within $[h_{min},h_{max}]$ and the fact that for the periodic morphology-similar signals, there must be a complete pattern included in arbitrary segment with $h_{max}$ length; (3) Utilize start indexes of every two consecutive segments $\overline{S}^{*}_{i}$ and $\overline{S}^{*}_{i+1}$ in overlapping matching segmentation $\overline{S}^{*}$ to reform the final non-overlapping segmentation  $\overline{S}^{\dag} = \{\overline{S}^{\dag}_{1},\overline{S}^{\dag}_{2},...,\overline{S}^{\dag}_{l}\}$ as follows:
\begin{equation}
\begin{aligned}
\overline{S}^{\dag}_{i} = \{S(\tau(\overline{S}^{*}_{i})), ...., S(\tau(\overline{S}^{*}_{i+1}))\} , \overline{S}^{*}_{i}\in \overline{S}^{*}, i = 1,2,...
\end{aligned}
\end{equation}
\par Once we get the segmentation $\overline{S}^{\dag}$ of sequence non-overlapping segments, the template updating can be stated as follows:
\begin{equation}
\begin{aligned}
\overline{T}^{\dag} = \mathop{\arg\min}_{T} \sum_{\overline{S}^{\dag}_{i}\in \overline{S}^{\dag}} {\rm DTW}(T,\overline{S}^{\dag}_{i}).  \\
\end{aligned}
\end{equation}
To further reduce computational complexity, during template updating process, we replace dynamic time warping with linear time warping and the optimization is equal to a weighted least squares problem with the following closed-form solution:
\begin{equation}
\begin{aligned}
\overline{T}^{\dag} =  \frac{1}{l} \sum_{\overline{S}^{\dag}_{i}\in \overline{S}^{\dag}} {\rm LTW}(\overline{S}^{\dag}_{i},h_{max}),  \\
\end{aligned}
\end{equation}
where ${\rm LTW}(\overline{S}^{\dag}_{i},h_{max})$ denotes the linear time warping process, which are realized by linear interpolation of segment $\overline{S}^{\dag}_{i}$ into length $h_{max}$. Then the final cardiac motion pattern
matching score can be calculated by following equation:\\


\begin{eqnarray}
{\rm P}(S) = \frac{1}{l}\sum_{\overline{S}^{\dag}_{i}\in \overline{S}^{\dag}} {\rm DTW}(\overline{T}^{\dag},\overline{S}^{\dag}_{i}).
\end{eqnarray}

\begin{figure}
\centering
\includegraphics[width=1\linewidth]{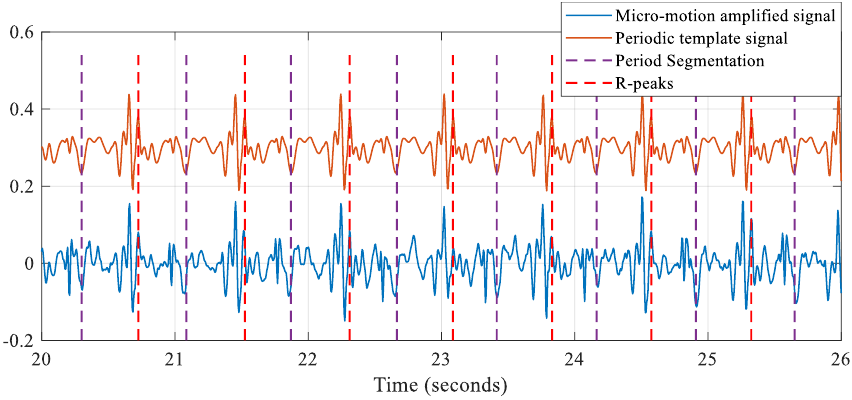}
\vspace{-0.4cm}
\caption{Result of pattern matching process.}
\label{fig2-c}
\vspace{-0.5cm}
\end{figure}

\par The Fig. \ref{fig2-c} shows the pattern matching result of micro-motion signals in a specific voxel. 
It can be seen that the cardiac motion pattern repeating along cardiac cycles is segmented and extracted. 
Also, the length of segmented cardiac cycle is synchronised with cardiac cycle as denoted by the ECG R-peaks. The signal is considered as the cardiac related, which has periodic pattern similar to the segmented signals. In contrast, the signal is identified as noise when the pattern matching process fails. The results of evaluation for all voxels are back projected to the 3D beamforming space as shown in Fig. \ref{fig2-a}(d). Note that the higher matching score voxels distributes around the chest and the scores decreases gradually with voxels away from the heart in distance. Finally, we only focus on the cardiac related reflections which are identified by valuable pattern matching process. Specifically, in our system we set a threshold $thr_{f}$ empirically, and the signals that have matching score less than $thr_{f}$ are eliminated for the subsequent process.  By aligning those spatial distributed micro-motion waveform in uniform time sequence, we analyze the morphology feature of the focused cardiac motion signals. 
From comparison result in Fig. \ref{fig2-a}(e), there are multiple morphology trends among all the waveform and each trend is surrounded by waveform with displacements variation. The result of these radar cardiac measurements difference both in spatial distribution and morphology variance are consistent with the mechanism of cardiac mechanical conduction. Therefore, it is clear to show the necessity and radar ability of measuring cardiac motion by observing the cardiac motion in a redundant sensing space including the entire thorax and its surrounding parts of torso.

\begin{figure*}[htbp]
\centering
\includegraphics[width=\linewidth,scale=1.00]{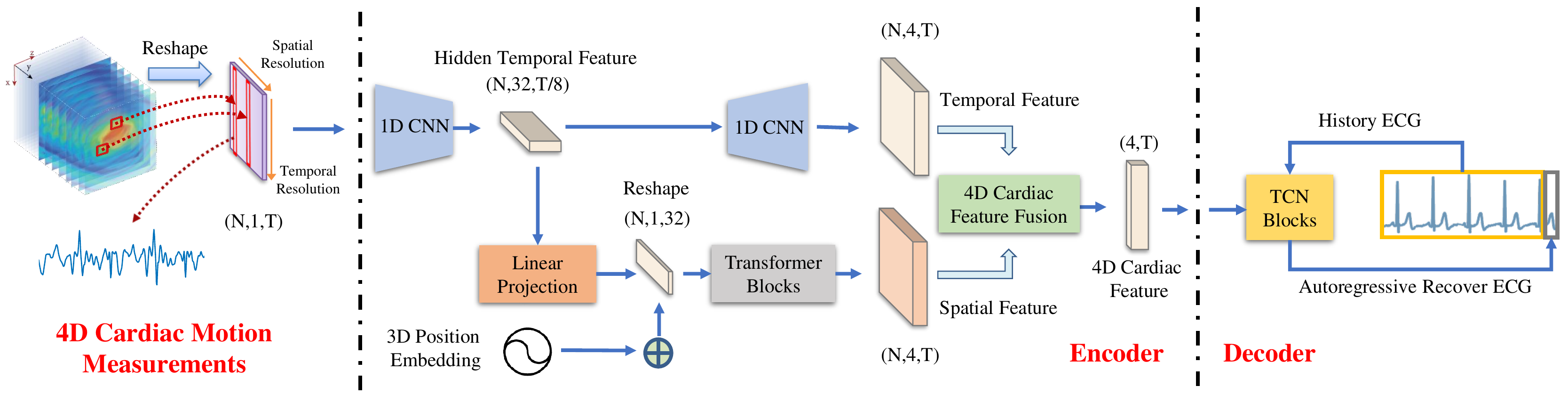}
\vspace{-0.4cm}
\caption{Deep neural network architecture for domain transformation.}
\label{fig3}
\vspace{-0.4cm}
\end{figure*}
\subsubsection{Spatial Filtering for Cardiac Signals}

\par The implementation of spatial filtering is intuitively come up with the fact that the cardiac signals are spread over the entire body surface and have similar motion trends with spatially nearby signals. Considering the spatial resolution limitation of radar sensing, the measurements sampled from the beamforming voxels are easily interfered by noise. By utilizing the spatial-similarity-based clustering, we can merge similar noisy signals to improve the robustness and SNR of the radar cardiac sensing statistically. We therefore develop a K-means clustering \cite{likas2003global} based spatial filter to merge and filter radar cardiac signals spatially. We perform K-means signal clustering within the constraints of position, reflected energy, and time-domain signal correlation. During filtering, the cardiac motion related signals around the 3D space are divided into pre-defined distinct non-overlapping clusters and the signals within clusters are merged to centroid. Specifically, we pre-define $N$ clusters and define the objective function for clustering as follows:
\begin{eqnarray}
\begin{aligned}
& \ \ \ \ \ \ J = \sum_{i=1}^{m}\sum_{k=1}^{K}(w_{i,k}\rho_{s}\|s_{i} - \mu_{k}\|^{2} + w_{i,k}\rho_{l}\|l_{i} - l_{\mu_{k}}\|^{2}), \\
& \ \ \ \ \ \ \ \ \mu_{k} = \frac{\sum_{i=1}^{m}w_{i,k}p_{i}s_{i}}{\sum_{i=1}^{m}w_{i,k}p_{i}}, \quad\quad  l_{\mu_{k}} = \frac{\sum_{i=1}^{m}w_{i,k}p_{i}l_{i}}{\sum_{i=1}^{m}w_{i,k}p_{i}},
\end{aligned}
\end{eqnarray}
where $w_{i,k} = 1$ for cardiac motion $s_{i}$ if it belongs to cluster $k$, otherwise, $w_{i,k} = 0$. $\mu_{k}$ is the centroid of cluster $k$, $l_{i}$ is the 3D location of the motion $s_{i}$, $l_{\mu_{k}}$ is the 3D location of cluster $k$ centroid. The objective function consists of the distance of motion difference $\|s_{i} - \mu_{k}\|^{2}$ and the distance of position difference $\|l_{i} - l_{\mu_{k}}\|^{2}$, $\rho_{s}$ and $\rho_{p}$ are balanced ratio for distance of motion difference and position difference respectively. We use common Expectation-Maximization(EM) method to solve the optimization problem. The cluster centroid $\mu_{k}$ and location of centroid $l_{\mu_{k}}$ are updated by the signal power $p_{i}$ weighted normalizing of motion signals $s_{i} $ and corresponding positions $l_{i}$ within the cluster respectively.

As shown in Fig. \ref{fig2-a}(f), the focused signals are further merged to the cluster center. With the spatial filtering, the signals in the same cluster are merged to promote the SNR and robustness of the radar sensing.  The Fig. \ref{fig2-a}(g) shows that the spatially filtered signal is more distinguishable in the time-domain representation, unlike the previous step where many neighboring signals gather together in time-domain. Also the compressed spatial redundancy has been utilized to improve the accuracy of motion measurements in radar. The motion signals are obtained mainly based on the phase extraction from the quadrature I/Q signals. However, this non-linear operation causes the information of reflection power lost, which is a reliable indicator of signal quality for radar measurements. For signals within a cluster, the power weighted spatial filter help us focusing on the signal with high SNR and suppress the noise interference. 
\par Finally, the 4D cardiac motion measurements are extracted after this sequence of signal processing algorithms, which can be expressed in the representation as $C_{S} = \{\mu_{n},l_{\mu_{n}}\}, \forall n \in 1,2,...,N$, where $\mu_{n}$ is a cardiac motion measurement sequence with k frames respect to the 3D location $l_{\mu_{n}}$. 

\subsection{Deep neural network for Domain Transformation}

\par The application of deep neural network 
has found widespread to the discipline of biomedical engineering. Most of those works are driven by human cognition purpose, such as medical image analysis \cite{shen2019patient}, disease diagnosis \cite{poplin2018prediction} and even predicting the effects of gene expression \cite{he2020integrating}, which have demonstrated the advantage of allowing real world data to guide cognitive representation through deep-learning training process. Considering the relation between mechanical conduction and electrical conduction from the cardiac cycle, we hypothesize that they share the same hidden feature, which are encoded separately from the same cardiac activity. Therefore, we extend the boundary of deep-learning to the cardiac related domains transformation and design a hierarchical deep-neural network that learns the reconstruction between cardiac electrical conduction and cardiac mechanical conduction, achieves end-to-end reconstruction mapping from RF input to the ECG output.

\par Fig. \ref{fig3} shows the detailed structure of our deep-learning network. Specifically, our deep-learning architecture based on encoder-decoder framework consists of 2 main parts: cardiac temporal-spatial features encoder, ECG reconstruction decoder. The input to the neural network is the 4D cardiac motion measurements. The output of the network is the corresponding ECG measurements. During the model training process, the network learns the mapping functions from RF signals to cardiac hidden feature and hidden feature back to the ECG separately.  

\par Cardiac cycle induced motions are driven by timing events of heart beat and its corresponding mechanical conduction in torso space jointly. Therefore, the underlying relation between the hidden feature representations and input signal data need to be encoded across all 4 dimensions (3D space and time dimension). We employ a hybrid CNN \cite{lecun1989backpropagation}-Transformer \cite{vaswani2017attention} architecture of encoder to extract both temporal and spatial features. The 1D CNN based encoder extracts the embedding features along time dimension, compresses the disorganized time-series information into high-level time-dependent abstract representation and learns a semantic temporal feature map for every signal in voxels. The following transformer blocks encode those spatially distributed temporal features and their corresponding spatial information from relative location collectively. With the self-attention mechanism and multi-head layer design, all the global and local spatial relation between voxels in 3D space could be extracted and represented in spatial feature map. Besides, by deep-learning the dynamic contribution of the spatial features to the final reconstruction task, the transformer blocks also help us 
fuse spatial feature map and temporal feature map jointly into the 4D cardiac feature $h$.

\par Since the cardiac cycle consists of cardiac events followed by the time order, the current cardiac state is only determined by prior information from previous states and the current cardiac related information. Correspondingly, given the encoded hidden feature $h_{t}$ at time $t$, the ECG reconstruction process is formulated as estimating the conditional distribution $p(X \mid h)$ as follows:
\begin{eqnarray}
\begin{aligned}
p(X \mid h) = \prod_{t=1}^{T}p(x_{t}\mid x_{1},...,x_{t-1},h_{t}).
\end{aligned}
\end{eqnarray}
Each ECG measurement $x_{t}$ is therefore conditioned on the sequence samples at previous timesteps and current hidden feature $h_{t}$. Motivated by the TCN \cite{lea2016temporal}\cite{bai2018empirical}, we model this conditional distribution with a stack of dilated convolutional blocks that receives $X_{1:T}$ and $h_{t}$ as inputs and outputs a distribution over possible $x_{t}$. All the ECG measurements are then computed in one forward sequence to sequence (Seq2Seq) pass. By using dilated casual convolutions, we make sure the model cannot violate the ordering in which we formulate the reconstruction process: the reconstruction emitted by the model at timestep $t$ cannot depend on any of the future timesteps. Also, the dilation factors increase exponentially to ensure a sufficiently large temporal context window to take advantage of the long-range dependencies of the ECG data. Considering the ECG data with 200Hz sample frequency, 12 TCN blocks stacking could provide us at least 20 seconds (4096 samples) receptive filed with better control of the model's memory size. However, in this case of long input sequence, other recurrent network architecture designed for Seq2Seq problem, such as LSTMs and GRUs, can easily use up a lot of memory to store the partial results for their multiple cell gates \cite{bai2018empirical}. At training time, the conditional reconstructions for multiple timesteps can be made in parallel since the same convolution filter is used in each layer, instead of sequentially as in other recurrent architectures. Also its back propagation path is different from the temporal direction of the sequence avoiding the problem of exploding/vanishing gradients.

\par The deep neural network is trained based on data-driven method by minimizing the error between the reconstructed ECG and the ground truth ECG.  As shown in Fig. \ref{fig3-a}(a), there are five consecutive feature waves named P, Q, R, S, T, which represent the electrical depolarization and repolarization throughout the cardiac cycle. Characteristic ECG tracing is defined by the variation of amplitude and interval respect to these waves. However the amplitudes of these waves are different statistically. For example, R peak usually has a much larger amplitude than other waves, since the R wave represents the rapid depolarization of the right and left ventricles which have a large muscle mass compared to the atria. Considering these statistical amplitude difference under different cardiac activities, the common loss functions for autoregressive learning task such as mean absolute error play much attention to the large amplitude variation (i.e. R-peak variation) and ignore the small amplitude variation (i.e. P-peak variation). To avoid such uneven penalty during learning, we apply a $\mu$-law companding transformation \cite{recommendation1988pulse} to the groudtruth ECG data first, and then quantize it to 256 possible value:
\begin{figure}
\centering
\includegraphics[width=1\linewidth]{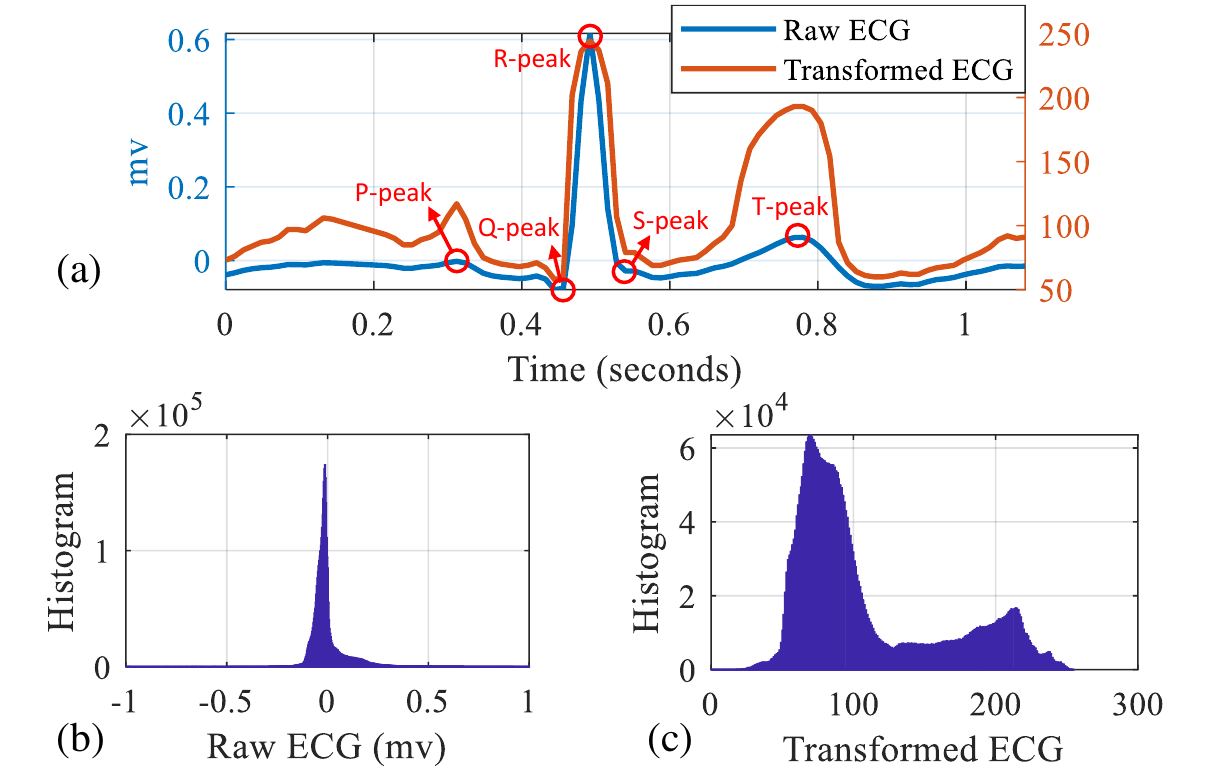}
\vspace{-0.3cm}
\caption{$\mu$-law transformation for ECG: (a) The comparison between the transformed ECG and the raw ECG with in a cardiac cycle; (b) The distribution of raw ECG; (c) The distribution of transformed ECG.}
\label{fig3-a}
\vspace{-0.3cm}
\end{figure}
\begin{eqnarray}
\begin{aligned}
f(x_{t}) = {\rm sign}(x_{t})\frac{\ln(1+\mu|x_{t}|)}{\ln{(1+\mu)}},
\end{aligned}
\end{eqnarray}
where $x_{t}$ is ground truth ECG at time t and normalized in $-1 < x_{t} < 1$, $\mu = 255$. We show a result of $\mu$-law transformation of one cycle of ECG and the distribution difference between raw ECG and quantized ECG over the entire ECG dataset in Fig. \ref{fig3-a}. This non-linear quantization enlarges the ECG amplitude with small numerical changes, shrinks the amplitude with larger numerical changes and makes the amplitude distribution more even. We propose to use softmax distribution rather than the continuous numerical regression to represent the conditional distribution of the next ECG value and choose cross-entropy as the final loss function. The reasons is that a categorical distribution is more flexible and can more easily model arbitrary distributions without assumptions about their shape.

\section{Experiments}
\subsection{Experimental Setup}
\begin{figure}[h]
\centering
\includegraphics[width=0.8\linewidth]{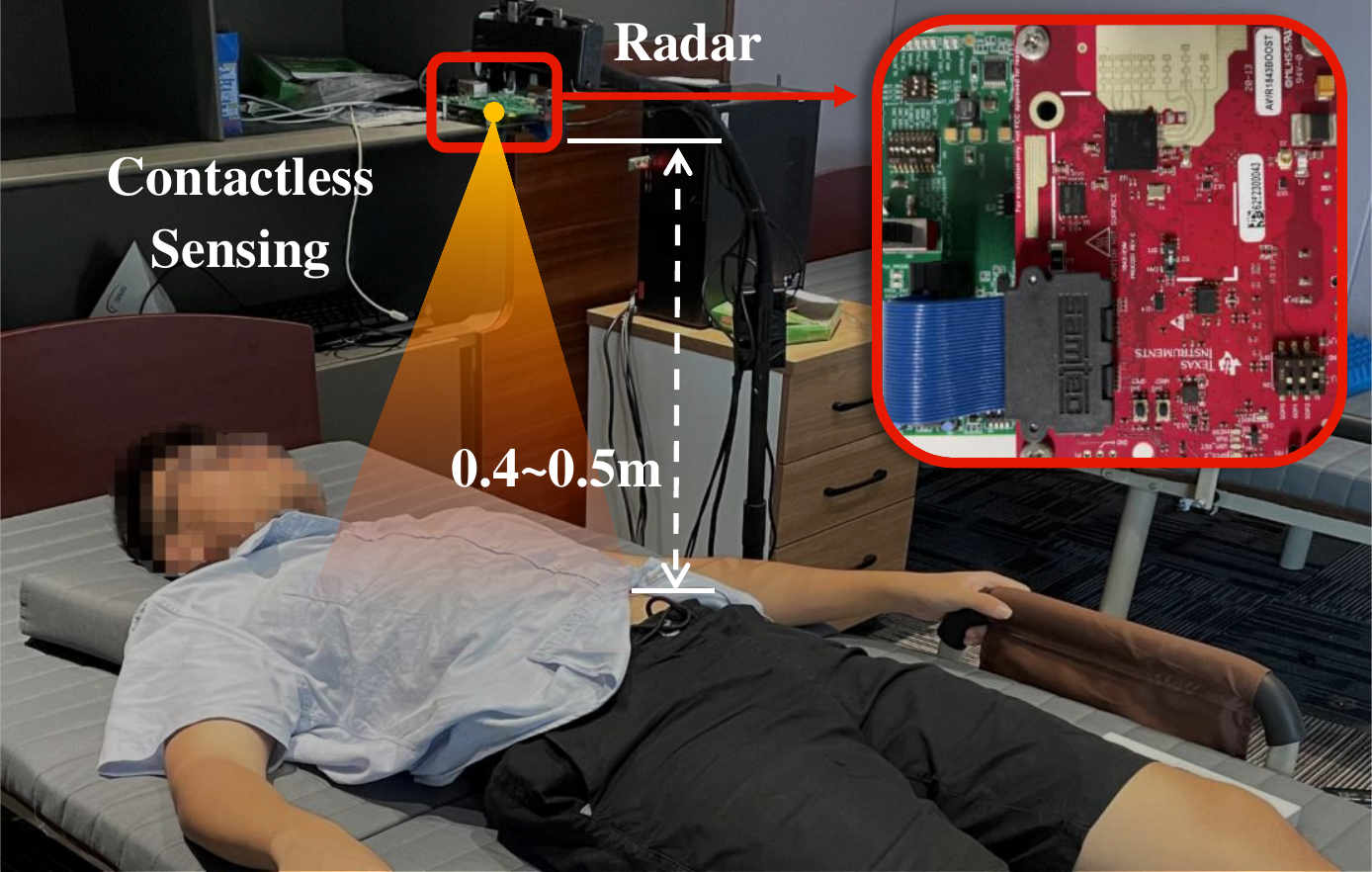}
\caption{Experimental setting for contactless ECG monitoring.}
\label{fig4-a}
\end{figure}

Considering the common scenarios of ECG monitoring in hospital, we perform our contactless monitoring system in a clinically relevant setting as shown in Fig. \ref{fig4-a}. During data acquisition, the participants are asked to lie in the supine position and remain quasi-static status. The radar sensor is placed above the torso chest within 0.4-0.5m and the main lobe of the antennas is directed to the sternum approximately. The ground truth ECG measurements were collected simultaneously with the radar measurements by an ECG monitor (TI ADS1292 evaluation board). 
Since training a deep neural network model requires a large amount of data with balanced distribution, we conducted 200 experimental trials over 35 participants (22males and 14 females) between the ages of 18 and 65. The trials are designed consisting of 4 different physiological status:  normal-breath, irregular-breath, post-exercise (for instance, jumping jacks) and sleep to expand the diversity of cardiac rhythms (including arrhythmia, bradycardia, tachycardia, normal rhythm) in the datasets. Also, this experimental setting helps us evaluate the overall performance considering common physiological conditions in daily life. Each trial lasts for 3 minutes. Given the 200 Hz sampling frequency of cardiac motions in radar, the total dataset consists around 7200000 frames of radar measurements and its corresponding ECG ground truth (around 40000 cardiac cycles with 50-125 beats per minute (BPM) heart rate variance) in total. 

\vspace{-0.6cm}
\subsection{Implementations Details}
\subsubsection{Radar Configuration}
\par We implement our contactless ECG monitoring system by using TI AWR1843 millimeter-wave radar and DCA1000 real-time data acquisition board. We activate 3 transmitters (Tx) and 4 receivers (Rx) to achieve a virtual 2D antenna array with 12 channels. Time division multiplexing strategy is exploited to achieve signal orthogonal in time among multiple Tx antennas. During one frame of radar sensing, all the 3 Tx transmit chirps of RF signal successively with $45\mu s$ interval to acquire the baseband signal from one channel to the entire 4 Rxs. The detailed configuration of chirp and frame parameters are shown in Table \ref{tab_params}. Under these settings, the radar achieves a frame rate of 200Hz, total 3.32Ghz bandwidth.

\begin{table}[htbp]
\begin{center}
 \caption{Parameters setting of radar chirps and frame}
 \label{param}
 \begin{tabular}{ll}
  \toprule
  \textbf{Parameter}  & \textbf{Value}\\
  \midrule
  Start frequency & { 77GHz} \\
  Frequency slope & {65MHz/µs} \\
  Idle time & {10µs} \\
  Ramp end time & {60µs} \\
  Sample points & {256} \\
  Sample rate & {5MHz} \\
  Frame periodicity & {5ms} \\
  \bottomrule
 \end{tabular}%
 \label{tab_params}%
 \end{center}
\end{table}%
\vspace{-0.3cm}

\subsubsection{Signal Processing}
\par Radar signal processing algorithm is performed offline using Matlab. The 3D beamforming is conducted in a radar relative coordinate. For computational efficiency, we crop the 3D sensing grid into a size of $-0.4$m$<x<0.4$m, $-0.4$m$<y<0.4$m, $0.35$m$<z<0.6$m, which are quantified and represented in a 9$\times$17$\times$9 = 1377 voxels. The x, y, z axis are parallel to the direction of body height, body width, and vertical to chest approximately. The cropping space is redundant for the cardiac motion measurements considering the distance between the radar and the torso as well as the size of the human body in our experimental setting. The centroid number N of K-means clustering is set as 50.

\subsubsection{Network Implementation}
\par The first CNN network includes 4 repeated layers of two 1D convolutions with kernal size of 7 (padded convolutions). Each convolution is followed by a rectified linear unit (ReLU) and batch normalization. The output from the layers follows a 1$\times$2 max-pooling operation with stride 2 for down sampling. 
In our implementation, we set the time sample length $l=640$. Considering the N=50 output signals of the signal processing, the input data size is $50 \times 1 \times640$ and the temporal encoder output size is $50 \times 32 \times 80$. The spatial encoder includes 3 blocks of transformer. At first, the input temporal feature are linear projected to the size of transformer input with $50 \times 32$. And the corresponding paired 3D position information are embedded linearly: $50 \times 3 \to 50 \times 32$. We set the transformer block with 4 heads of attention, dimension of Q/K/V = 64, feed-forward dimension of 128. The final extracted spatial features have size with $50 \times 32$. To bridge the encoder and decoder networks, we expand the dimension of both temporal and spatial features. As for the temporal features, the expansion is followed by the second CNN which consists of 4 layers of convolutions. And each layer is started with an up-sampling by the transposed convolution and two convolutions with kernal of 7 that halve the number of features. With a temporal feature of $50 \times 32 \times 80$, the data flow of the temporal feature maps leads to final dimension of $50 \times 4\times 640$. As for spatial feature, linear projection is utilized to expand the dimension same with: $50 \times 4\times 640$. The final cardiac features are fused by dot producting between expanded temporal feature and spatial feature. Then the size of cardiac feature is reshaped to $4 \times 640$ by linear projecting. We deploy a 9 stacks of TCN with dilation factor of 2, which provides a receptive field of 512.  During training, the ECG reconstructions can be made in parallel with step equals to 128 (input time series size $-$ receptive filed size), while during interference, the ECG reconstructions are made in autoregressive way with single step.

\begin{figure*}[htbp]
\centering
\includegraphics[width=\linewidth,scale=1.00]{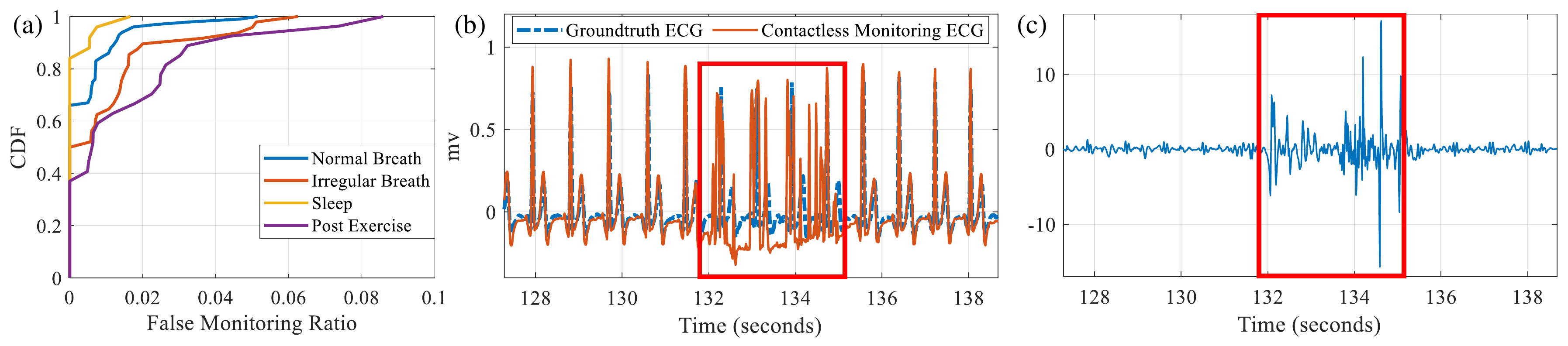}
\vspace{-0.4cm}
\caption{Overrall performance analysis: (a) CDF of false monitoring ratio for different physiological status; (b) Failed ECG reconstruction caused by body movement. The period of movement is marked by the red box; (c) Simultaneous radar cardiac motion measurements respect to a voxel.}
\label{fig4-b}
\vspace{-0.4cm}
\end{figure*}

\par We implement the network by using the PyTorch library \cite{paszke2017automatic}, and use the Adam optimizer \cite{kingma2014adam} to minimize the loss function and to update the network parameters iteratively. A learning rate of 0.001 and a mini-batch size of 64 are used. Data samples are segmented in length of 640 with sample step = 30. Considering the 200Hz sample frequency, every 3 minutes data provides 1100 training samples.

\vspace{-0.3cm}
\subsection{System Performance}
\par Our system is expected to have integral ECG monitoring potential. Therefore, we need to analyze the system ability in the scope of contactless ECG. In practical ECG diagnosis, any deviation from the normal tracing is potentially pathological and therefore of clinical significance. In this study, we focus on the quantifiable accuracy of cardiac events timing (Q, R, S, T waves) and waveform morphology, which are basics in ECG diagnosis and also provide clinically important information.
\par The accuracy of events timing in reconstructed ECG is determined by comparing the delineation between the ground truth and reconstructed ECG, which includes the identification and timing of those consecutive waves. To guarantee the independence and objective during this evaluation, we delineate both the ground-truth and reconstructed ECG by using Neurokit2 \cite{Makowski2021neurokit}, an open source package for ECG process routine. To quantify the accuracy of morphology between reconstructed ECG and the ground truth, we compute the Pearson-Correlation and Root-Mean-Square-Error (RMSE). Specifically, we calculate the correlation and RMS with respect to the reconstructed waveform segmented by the ground truth heart beat cycles.
\par Moreover, it is important to investigate the system performance under unseen patients. To do so, we divide our dataset into training set and testing set. For each participant, the model is trained on 34 other participants and evaluated on the untrained participant. This 35-fold cross validation ensures that the training and testing are mutually exclusive.

\subsubsection{Overrall Performance}

\par We first evaluate the system capability for monitoring ECG. According to the recommendations of the Association for Medical Instrumentation, it is considered that an onset, peak or an offset are detected correctly, if their deviation from the true annotations does not exceed in the absolute tolerance value of 150 ms \cite{test}. 
Therefore, we define the heart beat intervals in reconstructed ECG has correct monitoring result, if the consecutive waves within the beat cycle have clear waveform representation of onset, offset, peak and could been delineated by the Neurokit within 150 ms time error compared to the groundtruth ECG. Otherwise, the intervals are considered as false monitoring results. We calculate the ratio of number of heart beats which have false result to the total number of heart beats for every 3 minutes trials.
Fig. \ref{fig4-b}(a) plots the cumulative distribution function (CDF) of the false monitoring ratio for 4 physiological status. It is shown that the system can achieve $0.5\%$, $1.2\%$, $2\%$, $3.5\%$ 90-percentile false monitoring ratio during sleep, normal-breath, irregular-breath, post-exercise respectively. We recall those failed results and find out that the source of false results is due to interference from other random body motions which degrades the performance of the radar sensing of cardiac micro-motion and finally leads to the meaningless result. Fig. \ref{fig4-b}(b) shows the reconstruction result when the participant shakes his body. As shown in Fig. \ref{fig4-b}(c), during the torso shaking, the radar measurements are corrupted, which also leads to the messy ECG reconstruction result. It can be seen that the radar measurements return to stable state while the participant keeps quasi-static again and followed by the reasonable ECG reconstruction. This also explains the difference in the CDF plot, since the normal-breath, irregular-breath and post-exercise bring more and more random body motions compared to the static sleep state.

\subsubsection{Timing Accuracy for Cardiac Events}
\par
\begin{figure}
\centering
\includegraphics[width=\linewidth]{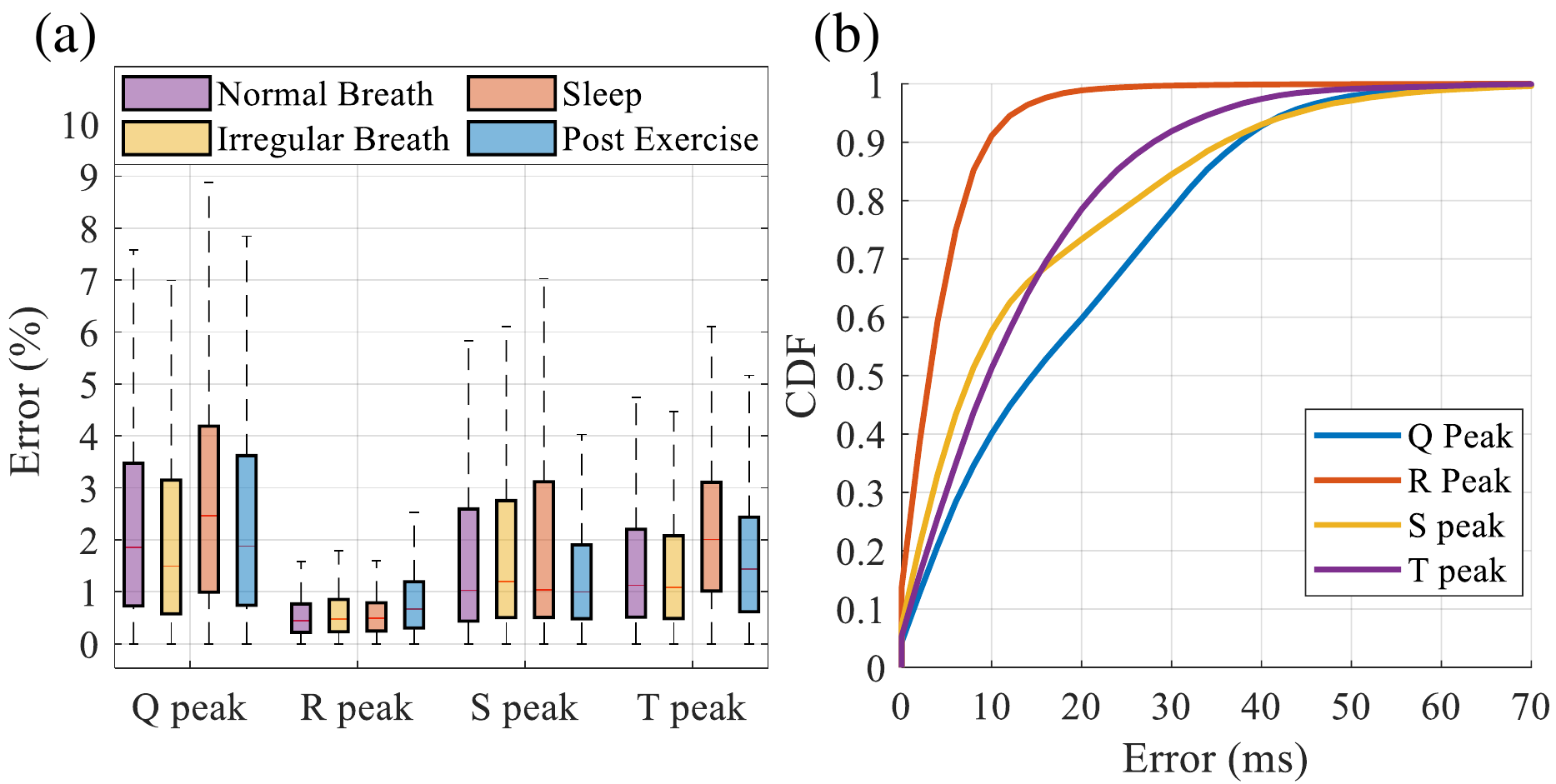}
\vspace{-0.4cm}
\caption{Timing accuracy analysis: (a) ECG events normalized timing error across 4 physiological status; (b) The CDF of the overall absolute timing error for the ECG events.}
\label{fig4-c}
\vspace{-0.4cm}
\end{figure}

 Based on the reconstructed ECG, we further evaluate the precision of ECG events timing. We normalize the events timing error by computing ratio of the error as the time difference between the reconstructed ECG timing and the groundtruth ECG timing to the heartbeat period. Fig. \ref{fig4-c}(a) shows ECG events timing error cross 4 physiological status. Each cluster represents one of the component waves peak and their corresponding statistic timing accuracy. The median normalized error of Q peak during normal-breath, irregular-breath, sleep, post-exercise status are observed as 1.85$\%$, 1.5$\%$, 2.4$\%$, 1.9$\%$, respectively. The median normalized error of R peak are 0.4$\%$, 0.5$\%$, 0.5$\%$, 0.7$\%$, respectively. The median normalized error of S peak are 1$\%$, 1.2$\%$, 1$\%$, 1$\%$, respectively. The median normalized error of T peak are 1.1$\%$, 1$\%$, 2$\%$, 1.5$\%$, respectively. We note that timing error during 4 status remain closely respect to the Q, R, S, T peak, which demonstrates that our system can achieve consistent timing performance with different chest motion interference. We also observe that the timing of R peak is better than that for the T peak, while both R peak and T peak are better for Q peak and S peak. 
The reason for such performance difference is that the discrepancy of amplitudes distribution among those peaks still exist even the $\mu$-law transformation is applied leading the neural network pay more attention to the larger amplitude variation (i.e. R peaks). To thoroughly evaluate the delineation accuracy in reconstructed ECG, Fig. \ref{fig4-c}(b) plots the CDF of the overall absolute timing error for the 4 peaks. Each of the CDFs in the figure corresponds to the combined error across all subjects and all physiological status. The median error in absolute timing are: 14ms (Q-peak), 3ms (R-peak), 8ms (S-peak), 10ms (T peak). The 90-percentile error in absolute timing is 38ms (Q-peak),  9ms (R-peak), 36ms (S-peak), 28ms  (T-peak). Similar to our earlier analysis, our system has higher accuracy in timing R peak compared to the T peak, Q peak and S peak. It is worth noticing that the best-case accuracy is limited by the temporal resolution of the radar cardiac sensing which results from its sampling period (200Hz/5 ms). We make the following remarks:
\begin{itemize}
    \item[$\bullet$]  Beyond these numerical accuracies of cardiac events timing, the results provide further implications in the scope of clinical diagnosis.  For example, the R-R interval analysis is the golden standard for diagnosing heart arrhythmia and an abnormally Q-T interval is associated with abnormal heart rhythms and sudden cardiac death. The above clinical indicators are calculated by corresponding cardiac events timing directly. 
    
    \item[$\bullet$] Considering that the loss function of the network implementation focuses on the categorical distribution of the predicted ECG values with no emphasis penalty on the timing accuracy, it demonstrates that our data-driven domain transformation design could find the relation mapping since the results reveal the meaningful information of cardiac electrical activities rather than the cycling waveform variation. Therefore, the performance difference over Q, R, S, T timing could probably be solved by a timing-accuracy-oriented loss function design in our future research. For example, we can decompose the ECG groundtruth to sub-waveforms that only contain single cardiac event morphology and calculate the loss separately to avoid the discrepancy of amplitudes distribution among those event peaks.

 \end{itemize}

\begin{figure}[h!]
\centering
\includegraphics[width=\linewidth]{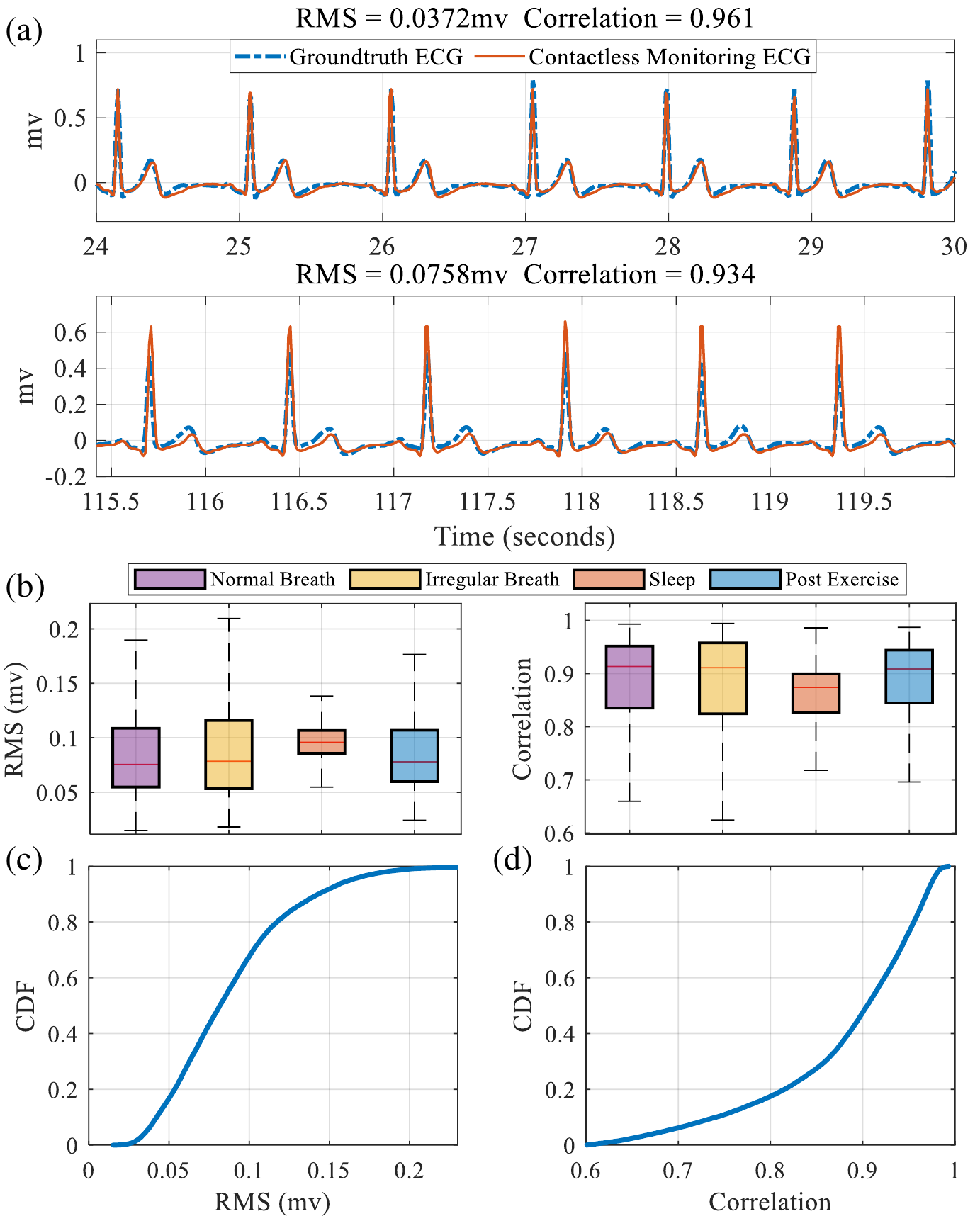}
\vspace{-0.3cm}
\caption{Morphology accuracy analysis: (a) Examples of contactless ECG monitoring result; (b) Correlation and RMS error across 4 physiological status; (c) The CDF of the overall RMS error; (d) The CDF of the overall correlation.}
\label{fig4-d}
\vspace{-0.4cm}
\end{figure}

\begin{figure*}[htbp]
\centering
\includegraphics[width=\linewidth,scale=1.00]{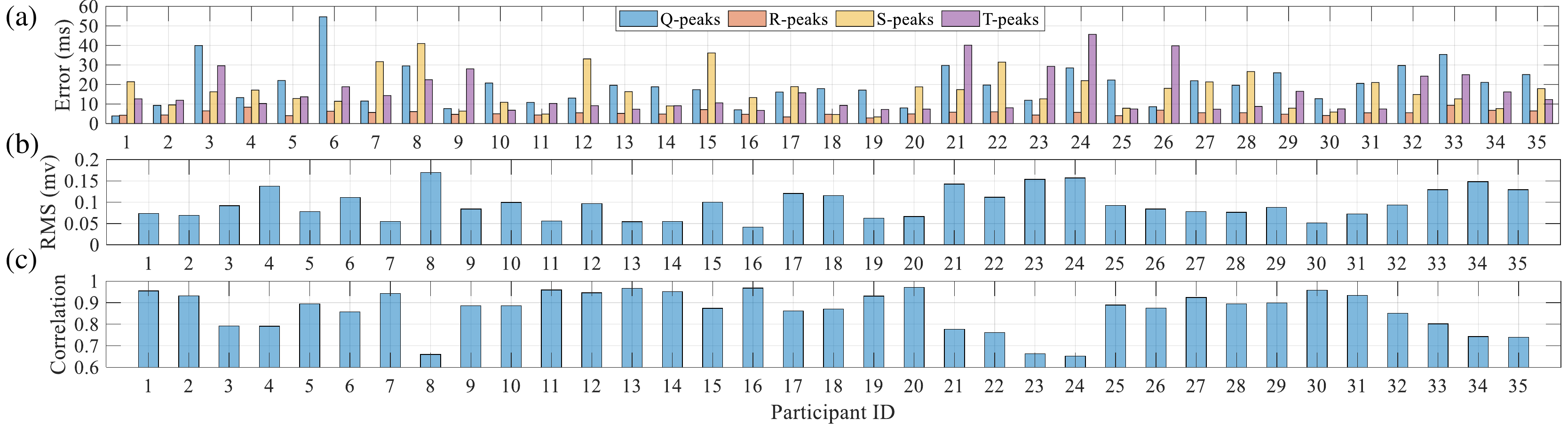}
\vspace{-0.4cm}
\caption{System performance across all participants: (a) Mean absolute error in timing each of the four feature points across different participants; (b) Mean RMS errors across participants; (c) Mean correlations across participants.} 
\label{fig4-e}
\vspace{-0.4cm}
\end{figure*}
\subsubsection{Morphology Accuracy}

\par Then, we analyze the morphology accuracy of the reconstructed ECG. In Fig. \ref{fig4-d}(a), we show two segments of contactless ECG monitoring results and the corresponding ground truth. It can be seen that the reconstructed ECG are basically same with the ground truth when its has low RMS error and high correlation. Fig. \ref{fig4-d}(b) shows the correlation and RMS accuracy between the reconstructed ECG and groundtruth ECG cross 4 physiological status. The system achieves median correlation of 91$\%$, 91$\%$, 87$\%$, 91$\%$ and median RMS error of 0.075mv, 0.079mv, 0.096mv, 0.078mv during normal breath, irregular breath, sleep and post exercise respectively. Similar to the result of the timing analysis, the morphology accuracy still remains closely, as expected to the consistent performance over different physiological status. 
Fig. \ref{fig4-d}(c) plots the CDF of the overall RMS error. The median RMS error is 0.081mv and the 90-percentile RMS error is 0.143mv. Fig. \ref{fig4-d}(d) plots the CDF of the overall correlation. The median correlation is $90\%$ and 90-percentile correlation is $74\%$. Thus, our contactless system provides contactless monitoring of the ECG with strongly correlation and low RMS error compared to the groundtruth. We make the following remark:
\begin{itemize}
    \item[$\bullet$]  Cardiac activity is rather complex. The further interpretation and diagnosis of ECG morphology are sophisticated and mainly are achieved subjectively by cardiologists. For example, an unusually tall QRS complex may represent left ventricular hypertrophy while a very low-amplitude QRS complex may represent a pericardial effusion or myocardial disease. Although the proposed system is evaluated concerning the fundamental accuracy of morphology, the evaluation in this work is not comprehensive in the clinical application scope. As the research evolves, it would be interesting to evaluate the system's applicable range to clinical diagnosis focusing on cardiac disease patients.

 \end{itemize}

\subsubsection{Clinical Usage Potential Analysis}

\par After analyzing the timing and morphology accuracy, it is important to demonstrate the potential of clinical usage for our contactlesss ECG monitoring system. We first evaluate the system performance for unseen participants based on our 35-fold cross-validation strategy. In Fig. \ref{fig4-e}, we plot the mean value of timing and morphology accuracy across 35 untrained participants. Fig. \ref{fig4-e}(a) shows that system achieves timing accuracy between 4ms-54ms, 3ms-9ms, 3ms-40ms, 6ms-45ms for Q, R, S, T peaks respectively across all participants. In Fig.  \ref{fig4-e}(b)(c), we demonstrate the system ability of monitoring ECG with RMS error between 0.04mv-0.17mv and correlation between 65\%-97\%. Similar to our earlier result, the system has better accuracy in timing R peak than other peaks. Generally, the system can maintain timing accuracy bellow 54ms and RMS error bellow 0.17mv across all participants and high correlation ($>0.8$) for 26 out of 35 participants. It is worth noticing that the timing and morphology accuracy have same variation trend respect to the participant. For example, for the 11th participant, system exceeds the overall performance with timing accuracy in 11ms, 4ms, 5ms, 10ms for Q, R, S, T peaks and morphology accuracy in RMS error of 0.05mv and correlation of 95\%. However, the 24th participant has relatively degraded performance with coordinated timing and morphology accuracy decrease 28ms, 6ms, 22ms, 45ms for Q, R, S, T peaks and morphology accuracy in RMS error of 0.16mv and correlation of 0.65. This demonstrates that our network design has consistent interpretability in accuracy, although the final loss function does not emphasize them.
\par
\begin{figure}
\centering
\includegraphics[width=0.95\linewidth]{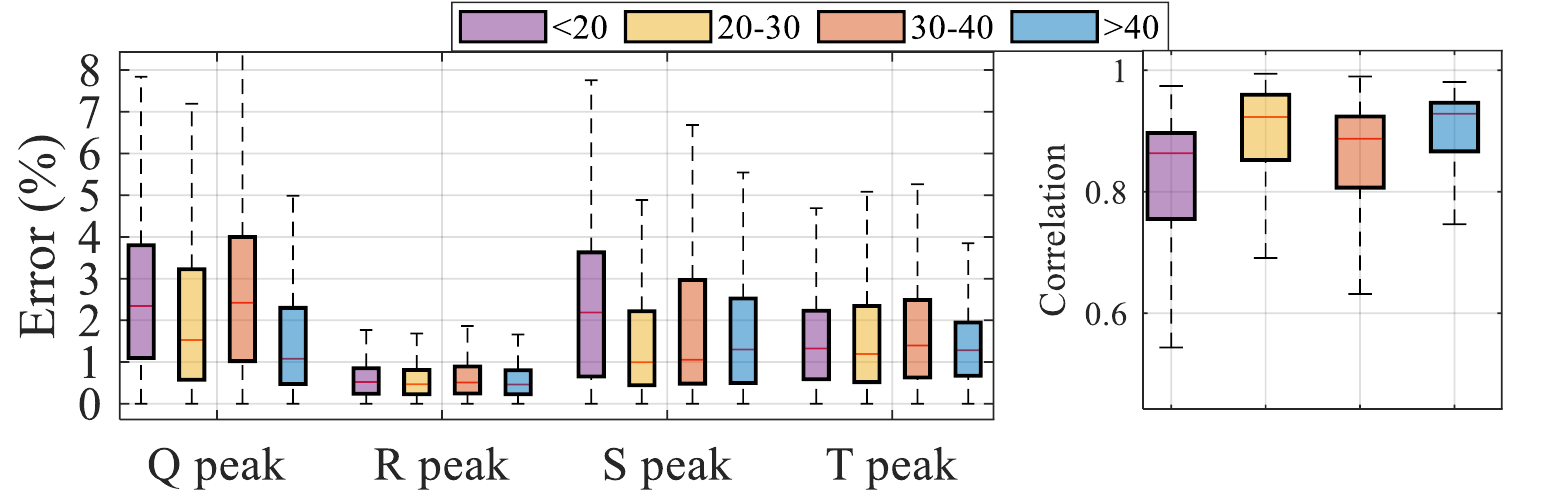}
\caption{System performance with 4 different age groups.}
\label{fig5}
\vspace{-0.5cm}
\end{figure}

 Also, our contactless ECG system is expected to have no bias on different age groups. Therefore, we divide the participants into four age groups and the system performance is shown in Fig.\ref{fig5}. The median normalized error of Q peak across 4 age groups are observed as 2.3$\%$, 1.5$\%$, 2.5$\%$, 1.0$\%$, respectively. The median normalized error of R peak are 0.5$\%$, 0.5$\%$, 0.5$\%$, 0.5$\%$, respectively. The median normalized error of S peak are 2.1$\%$, 1.0$\%$, 1.0$\%$, 1.2$\%$, respectively. The median normalized error of T peak are 1.3$\%$, 1.2$\%$, 1.4$\%$, 1.3$\%$, respectively. The median of correlation are 86$\%$, 92$\%$, 89$\%$, 93$\%$, respectively. We note that timing error and morphology accuracy across 4 age groups remain closely, which demonstrates that our system can achieve consistent performance with different ages.

\begin{figure}
\centering
\includegraphics[width=0.95\linewidth]{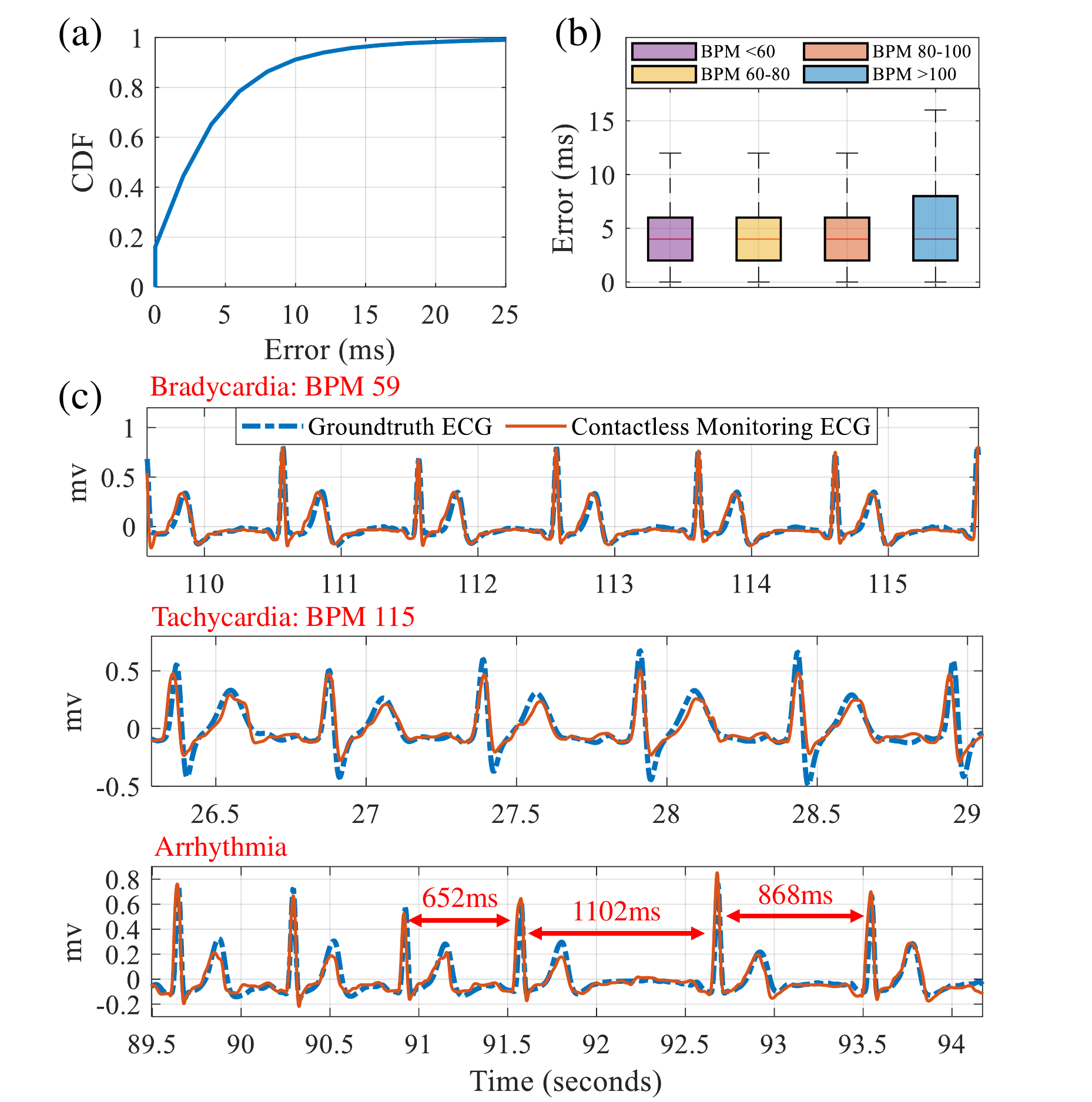}
\caption{Arrhythmias diagnosis potential analysis: (a) The CDF of the overall absolute timing error for R-R interval timing; (b) The absolute timing error of R-R interval timing under different heart rate conditions; (c) ECG reconstruction results under the condition of bradycardia, tachycardia, and arrhythmia.}
\label{fig4-f}
\vspace{-0.7cm}
\end{figure}
\par To further prove the system potential for clinical scenarios, we analyze the heart arrhythmias diagnosis based on our system. Heart arrhythmias occur when the electrical signals that coordinate the heart's beats don't work properly. The faulty signaling causes the tachycardia, bradycardia or irregularly heatbeats. There are plenty of heart arrhythmia instances in our dataset, since the experimental setting of 4 different physiological status could expand the diversity of heatbeats and stimulate different kinds of arrhythmias. Considering that analysis of the heart rate variability has been proved to be a powerful tool to assess heart arrhythmia \cite{acharya2006heart}, we analyze the R-R interval accuracy of our result. Fig. \ref{fig4-f}(a) shows the CDF of absolute timing error for R-R interval over the entire dataset. The median error of timing R-R interval is 3ms and the 90-percentile error is 9ms.  We also investigate the timing accuracy under different heart rate conditions. As shown in Fig. \ref{fig4-f}(b), our system can achieve the consistent performance of timing R-R interval since the error distributions remain almost the same when BPM $<100$ and for BPM $>100$ condition, the system still has the same median timing error but with larger 75-percentile error of 8ms. In Fig. \ref{fig4-f}(c), we show the ECG reconstruction results under the condition of bradycardia, tachycardia, arrhythmia, respectively. The analysis above indicates great robustness and accuracy of timing R-R interval and clinical potential for arrhythmias diagnosis. We make the following remark:
\begin{itemize}
    \item[$\bullet$]  Limited by the resolution of the radar device, the proposed system could be affected by the interference. The existence of interference could lead to reflected signal distortion and finally result in the lower SNR of cardiac mechanical motion measurements. In such a case, the domain transformation would misunderstand the cardiac motion situation considering the motion signal distortion caused by interference and generate low accuracy results (“flattening” artifacts in Fig. \ref{fig4-f}(c)) or even meaningless results in Fig. \ref{fig4-b}(b) with interference levels increasing. It would be valuable to expand the system's capability in dealing with interference so that it can improve the accuracy and robustness of the system in our future research.

 \end{itemize}

\subsubsection{Daily Life Usage Analysis}
The system is expected to apply to daily life environment. Therefore, it is essential to analyze the system constriction considering the impact of interference from adjacent people’s activity and measurement distance. To do so, We conduct contactless ECG monitoring over five new participants (not in the database) in a new environment with variation of adjacent interference and measurement distance.
\par
\begin{figure}
\centering
\includegraphics[width=0.85\linewidth]{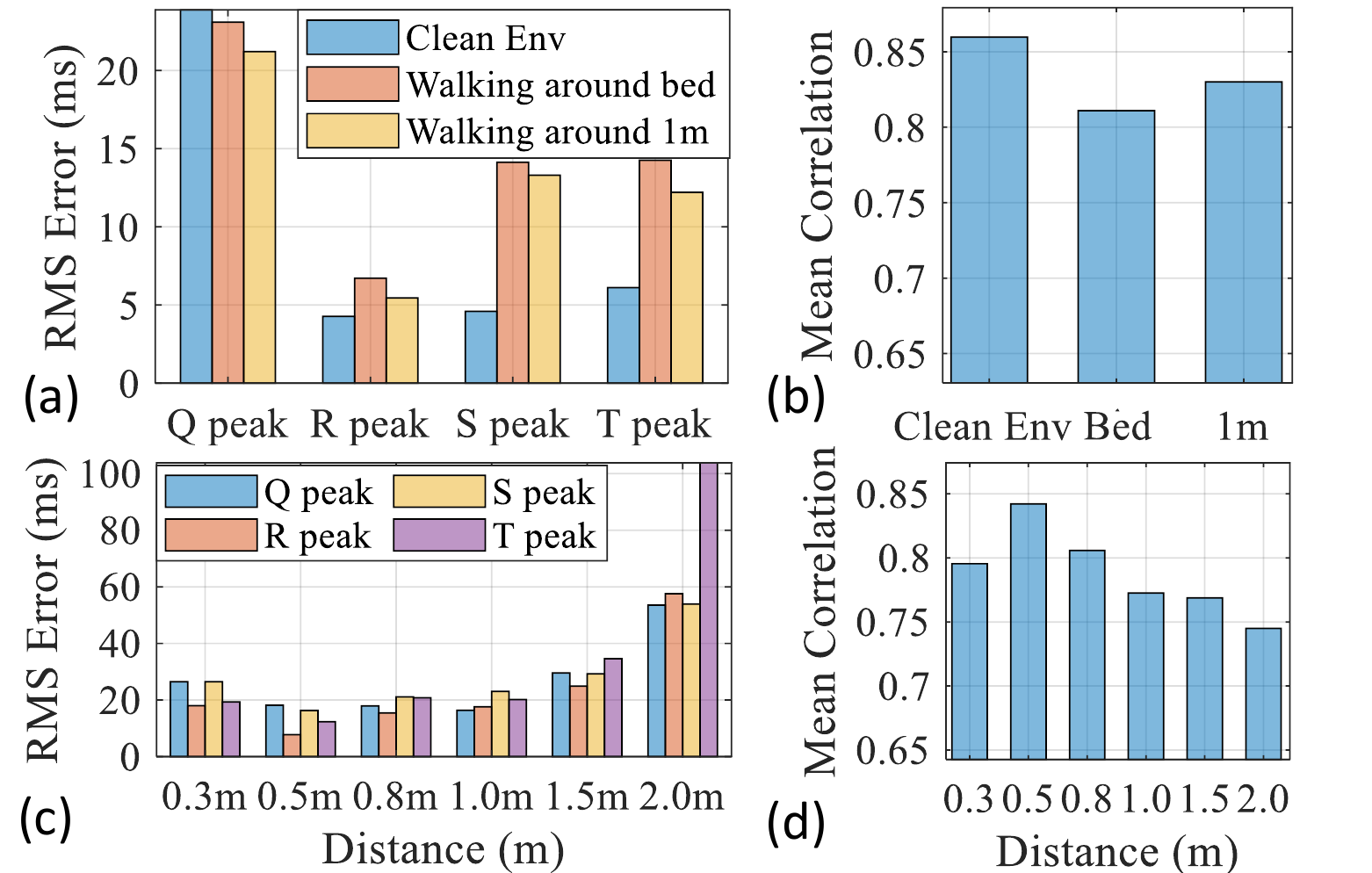}
\caption{Daily life usage analysis: (a) The RMS error of events timing under different interference levels; (b) The mean correlation under different interference levels; (c) The RMS error of events timing with measurement distance variation; (d) The mean correlation with measurement distance variation;} 
\label{fig6}
\vspace{-0.4cm}
\end{figure}

 We first examine the impact of adjacent interference. In this experiment, monitoring is implemented under situation of clean enviornment, people walking closely around the bed, and people walking around the bed 1 meter away to stimulate different interference level. As shown in Fig. \ref{fig6}(a)(b), the best performance is acquired in the clean environment. The performance of R, S, T timing and correlation deteriorate under the adjacent people’s interference. And the deterioration alleviates as the interference distance is away from the participant. 
\par Then, we examine the impact of measurement distance. We conduct experiments for measuring with different sensing distance. As shown in Fig. \ref{fig6}(c)(d), It is clear that the system achieves the best performance at around 0.5m. The performance decreases as the distance increases. The reason is that the path attenuation of RF signals and the deterioration of spatial resolution decreases the SNR of cardiac mechanical measurements. Also, it is interesting that the system performance decreases as the distance decreases from 0.5m, which is inconsistent with the intuition that closer radar sensing leads to higher SNR measurements. Such a phenomenon is due to the following reason. According to \cite{tiref}, the AWR1843 radar has 6dB-beamwidth of $\pm50$ and $\pm 20$ degrees in horizontal and vertical respectively, and the reflections outside from the beamwidth space is hard to be sensed by the system. Then, it can be derived that at 0.3m distance the radar only could capture $20\times70cm^{2}$ space, which is not sufficient to cover the entire torso surface. Therefore, the incomplete cardiac radar measurements also lead to the performance degradation.
\par According to the analysis above, the system achieves the best performance in a clean environment at around 0.5m sensing distance. However, depending on the performance requirements of a specific application, the system could work properly with a wide range of setting.\\

\vspace{-0.7cm}
\section{Conclusion}
\par We have demonstrated a novel ECG monitoring system that offered contactless, accurate and continuous monitoring of cardiac electrical activity from RF signal, adding a new dimension to the sensing range of radar. This system exploited high accuracy sensing of cardiac mechanical activity with well designed signal processing algorithms, learned cardiac related domain transformation through deep learning and reconstructed ECG output with RF input in an end-to-end manner. We provided the evidence of the feasibility and potential of such contactless ECG monitoring system through extensive experiments. Our evaluation in this paper has focused on healthy individuals. As the research evolves, it would be very interesting to evaluate the system’s accuracy in medical applications.

\bibliographystyle{IEEEtran}
\bibliography{references}

\begin{thebibliography}{10}
\providecommand{\url}[1]{#1}
\csname url@samestyle\endcsname
\providecommand{\newblock}{\relax}
\providecommand{\bibinfo}[2]{#2}
\providecommand{\BIBentrySTDinterwordspacing}{\spaceskip=0pt\relax}
\providecommand{\BIBentryALTinterwordstretchfactor}{4}
\providecommand{\BIBentryALTinterwordspacing}{\spaceskip=\fontdimen2\font plus
\BIBentryALTinterwordstretchfactor\fontdimen3\font minus
  \fontdimen4\font\relax}
\providecommand{\BIBforeignlanguage}[2]{{%
\expandafter\ifx\csname l@#1\endcsname\relax
\typeout{** WARNING: IEEEtran.bst: No hyphenation pattern has been}%
\typeout{** loaded for the language `#1'. Using the pattern for}%
\typeout{** the default language instead.}%
\else
\language=\csname l@#1\endcsname
\fi
#2}}
\providecommand{\BIBdecl}{\relax}
\BIBdecl

\bibitem{bib1}
G.~A. Roth, C.~O. Johnson, K.~H. Abate, F.~Abd-Allah, M.~Ahmed, K.~Alam,
  T.~Alam, N.~Alvis-Guzman, H.~Ansari, J.~{\"A}rnl{\"o}v \emph{et~al.}, ``The
  burden of cardiovascular diseases among us states, 1990-2016,'' \emph{JAMA
  cardiology}, vol.~3, no.~5, pp. 375--389, 2018.

\bibitem{bib2}
A.~Timmis, N.~Townsend, C.~Gale, R.~Grobbee, N.~Maniadakis, M.~Flather,
  E.~Wilkins, L.~Wright, R.~Vos, J.~Bax \emph{et~al.}, ``European society of
  cardiology: cardiovascular disease statistics 2017,'' \emph{European heart
  journal}, vol.~39, no.~7, pp. 508--579, 2018.

\bibitem{bib3}
{\"O}.~Yildirim, P.~Plawiak, R.~Tan, and U.~Acharya, ``Arrhythmia detection
  using deep convolutional neural network with long duration ecg signals,''
  \emph{Computers in biology and medicine}, vol. 102, pp. 411--420, 2018.

\bibitem{bib4}
M.~Bsoul, H.~Minn, and L.~Tamil, ``Apnea medassist: real-time sleep apnea
  monitor using single-lead ecg,'' \emph{IEEE transactions on information
  technology in biomedicine}, vol.~15, no.~3, pp. 416--427, 2010.

\bibitem{bib5}
S.~Preejith, R.~Dhinesh, J.~Joseph, and M.~Sivaprakasam, ``Wearable ecg
  platform for continuous cardiac monitoring,'' 2016, 2016 38th Annual
  International Conference of the IEEE Engineering in Medicine and Biology
  Society (EMBC).

\bibitem{bib6}
M.~M. Baig, H.~Gholamhosseini, and M.~J. Connolly, ``A comprehensive survey of
  wearable and wireless ecg monitoring systems for older adults,''
  \emph{Medical \& biological engineering \& computing}, vol.~51, no.~5, pp.
  485--495, 2013.

\bibitem{bib7}
M.~A. Serhani, H.~T~El~Kassabi, H.~Ismail, and A.~Nujum~Navaz, ``Ecg monitoring
  systems: Review, architecture, processes, and key challenges,''
  \emph{Sensors}, vol.~20, no.~6, p. 1796, 2020.

\bibitem{bib8}
D.~Zhang, Y.~Hu, and Y.~Chen, ``Mtrack: Tracking multiperson moving
  trajectories and vital signs with radio signals,'' \emph{IEEE Internet of
  Things Journal}, vol.~8, no.~5, pp. 3904--3914, 2020.

\bibitem{bib9}
M.~Mercuri, I.~R. Lorato, Y.-H. Liu, F.~Wieringa, C.~Van~Hoof, and T.~Torfs,
  ``Vital-sign monitoring and spatial tracking of multiple people using a
  contactless radar-based sensor,'' \emph{Nature Electronics}, vol.~2, no.~6,
  pp. 252--262, 2019.

\bibitem{bib10}
C.~Will, K.~Shi, S.~Schellenberger, T.~Steigleder, F.~Michler, J.~Fuchs,
  R.~Weigel, C.~Ostgathe, and A.~Koelpin, ``Radar-based heart sound
  detection,'' \emph{Scientific Reports}, vol.~8, 2018.

\bibitem{bib11}
C.~C. Moore, C.~H. Lugo-Olivieri, E.~R. McVeigh, and E.~A. Zerhouni,
  ``Three-dimensional systolic strain patterns in the normal human left
  ventricle: characterization with tagged mr imaging,'' \emph{Radiology}, vol.
  214, no.~2, pp. 453--466, 2000.

\bibitem{bib12}
Y.~Notomi, R.~M. Setser, T.~Shiota, M.~G. Martin-Miklovic, J.~A. Weaver, Z.~B.
  Popovic, H.~Yamada, N.~L. Greenberg, R.~D. White, and J.~D. Thomas,
  ``Assessment of left ventricular torsional deformation by doppler tissue
  imaging: validation study with tagged magnetic resonance imaging,''
  \emph{Circulation}, vol. 111, no.~9, pp. 1141--1147, 2005.

\bibitem{bib13}
B.~H. Amundsen, T.~Helle-Valle, T.~Edvardsen, H.~Torp, J.~Crosby, E.~Lyseggen,
  A.~St{\o}ylen, H.~Ihlen, J.~A. Lima, O.~A. Smiseth \emph{et~al.},
  ``Noninvasive myocardial strain measurement by speckle tracking
  echocardiography: validation against sonomicrometry and tagged magnetic
  resonance imaging,'' \emph{Journal of the American College of Cardiology},
  vol.~47, no.~4, pp. 789--793, 2006.

\bibitem{bib14}
M.~J. Tadi, E.~Lehtonen, A.~Saraste, J.~Tuominen, J.~Koskinen, M.~Ter{\"a}s,
  J.~Airaksinen, M.~P{\"a}nk{\"a}{\"a}l{\"a}, and T.~Koivisto,
  ``Gyrocardiography: A new non-invasive monitoring method for the assessment
  of cardiac mechanics and the estimation of hemodynamic variables,''
  \emph{Scientific reports}, vol.~7, no.~1, pp. 1--11, 2017.

\bibitem{bib15}
A.~Taebi, B.~E. Solar, A.~J. Bomar, R.~H. Sandler, and H.~A. Mansy, ``Recent
  advances in seismocardiography,'' \emph{Vibration}, vol.~2, no.~1, pp.
  64--86, 2019.

\bibitem{bib16}
C.~Fisch, ``Centennial of the string galvanometer and the electrocardiogram,''
  \emph{Journal of the American College of Cardiology}, vol.~36, no.~6, pp.
  1737--1745, 2000.

\bibitem{robey2004mimo}
F.~C. Robey, S.~Coutts, D.~Weikle, J.~C. McHarg, and K.~Cuomo, ``Mimo radar
  theory and experimental results,'' in \emph{Conference Record of the
  Thirty-Eighth Asilomar Conference on Signals, Systems and Computers, 2004.},
  vol.~1.\hskip 1em plus 0.5em minus 0.4em\relax IEEE, 2004, pp. 300--304.

\bibitem{ampanozi2018comparing}
G.~Ampanozi, E.~Krinke, P.~Laberke, W.~Schweitzer, M.~J. Thali, and L.~C.
  Ebert, ``Comparing fist size to heart size is not a viable technique to
  assess cardiomegaly,'' \emph{Cardiovascular Pathology}, vol.~36, pp. 1--5,
  2018.

\bibitem{drake2005gray}
R.~Drake, R.~L. Drake, H.~Gray, W.~Vogl, and A.~W. Mitchell, \emph{Gray's
  anatomy for students}.\hskip 1em plus 0.5em minus 0.4em\relax Elsevier Health
  Sciences TW, 2005.

\bibitem{owda2020reflectance}
A.~Y. Owda, N.~Salmon, A.~J. Casson, and M.~Owda, ``The reflectance of human
  skin in the millimeter-wave band,'' \emph{Sensors}, vol.~20, no.~5, p. 1480,
  2020.

\bibitem{de1997chest}
A.~De~Groote, M.~Wantier, G.~Ch{\'e}ron, M.~Estenne, and M.~Paiva, ``Chest wall
  motion during tidal breathing,'' \emph{Journal of Applied Physiology},
  vol.~83, no.~5, pp. 1531--1537, 1997.

\bibitem{ramachandran1989three}
G.~Ramachandran and M.~Singh, ``Three-dimensional reconstruction of cardiac
  displacement patterns on the chest wall during the p, qrs and t-segments of
  the ecg by laser speckle inteferometry,'' \emph{Medical and Biological
  Engineering and Computing}, vol.~27, no.~5, pp. 525--530, 1989.

\bibitem{wang2021mmhrv}
F.~Wang, X.~Zeng, C.~Wu, B.~Wang, and K.~R. Liu, ``mmhrv: Contactless heart
  rate variability monitoring using millimeter-wave radio,'' \emph{IEEE
  Internet of Things Journal}, 2021.

\bibitem{richards2014fundamentals}
M.~A. Richards, \emph{Fundamentals of radar signal processing}.\hskip 1em plus
  0.5em minus 0.4em\relax McGraw-Hill Education, 2014.

\bibitem{lanczos1996linear}
C.~Lanczos, \emph{Linear differential operators}.\hskip 1em plus 0.5em minus
  0.4em\relax SIAM, 1996.

\bibitem{lecun2015deep}
Y.~LeCun, Y.~Bengio, and G.~Hinton, ``Deep learning,'' \emph{nature}, vol. 521,
  no. 7553, pp. 436--444, 2015.

\bibitem{shen2019patient}
L.~Shen, W.~Zhao, and L.~Xing, ``Patient-specific reconstruction of volumetric
  computed tomography images from a single projection view via deep learning,''
  \emph{Nature biomedical engineering}, vol.~3, no.~11, pp. 880--888, 2019.

\bibitem{poplin2018prediction}
R.~Poplin, A.~V. Varadarajan, K.~Blumer, Y.~Liu, M.~V. McConnell, G.~S.
  Corrado, L.~Peng, and D.~R. Webster, ``Prediction of cardiovascular risk
  factors from retinal fundus photographs via deep learning,'' \emph{Nature
  Biomedical Engineering}, vol.~2, no.~3, pp. 158--164, 2018.

\bibitem{xiong2015human}
H.~Y. Xiong, B.~Alipanahi, L.~J. Lee, H.~Bretschneider, D.~Merico, R.~K. Yuen,
  Y.~Hua, S.~Gueroussov, H.~S. Najafabadi, T.~R. Hughes \emph{et~al.}, ``The
  human splicing code reveals new insights into the genetic determinants of
  disease,'' \emph{Science}, vol. 347, no. 6218, 2015.

\bibitem{he2020integrating}
B.~He, L.~Bergenstr{\aa}hle, L.~Stenbeck, A.~Abid, A.~Andersson, {\AA}.~Borg,
  J.~Maaskola, J.~Lundeberg, and J.~Zou, ``Integrating spatial gene expression
  and breast tumour morphology via deep learning,'' \emph{Nature biomedical
  engineering}, vol.~4, no.~8, pp. 827--834, 2020.

\bibitem{lea2016temporal}
C.~Lea, R.~Vidal, A.~Reiter, and G.~D. Hager, ``Temporal convolutional
  networks: A unified approach to action segmentation,'' in \emph{European
  Conference on Computer Vision}.\hskip 1em plus 0.5em minus 0.4em\relax
  Springer, 2016, pp. 47--54.

\bibitem{bai2018empirical}
S.~Bai, J.~Z. Kolter, and V.~Koltun, ``An empirical evaluation of generic
  convolutional and recurrent networks for sequence modeling,'' \emph{arXiv
  preprint arXiv:1803.01271}, 2018.

\bibitem{oord2016wavenet}
A.~v.~d. Oord, S.~Dieleman, H.~Zen, K.~Simonyan, O.~Vinyals, A.~Graves,
  N.~Kalchbrenner, A.~Senior, and K.~Kavukcuoglu, ``Wavenet: A generative model
  for raw audio,'' \emph{arXiv preprint arXiv:1609.03499}, 2016.

\bibitem{Makowski2021neurokit}
\BIBentryALTinterwordspacing
D.~Makowski, T.~Pham, Z.~J. Lau, J.~C. Brammer, F.~Lespinasse, H.~Pham,
  C.~Schölzel, and S.~H.~A. Chen, ``{NeuroKit}2: A python toolbox for
  neurophysiological signal processing,'' \emph{Behavior Research Methods},
  vol.~53, no.~4, pp. 1689--1696, feb 2021. [Online]. Available:
  \url{https://doi.org/10.3758%2Fs13428-020-01516-y}
\BIBentrySTDinterwordspacing

\bibitem{lin2016identification}
W.-Y. Lin, W.-C. Chou, P.-C. Chang, C.-C. Chou, M.-S. Wen, M.-Y. Ho, W.-C. Lee,
  M.-J. Hsieh, C.-C. Lin, T.-H. Tsai \emph{et~al.}, ``Identification of
  location specific feature points in a cardiac cycle using a novel
  seismocardiogram spectrum system,'' \emph{IEEE journal of biomedical and
  health informatics}, vol.~22, no.~2, pp. 442--449, 2016.

\bibitem{acharya2006heart}
U.~R. Acharya, K.~P. Joseph, N.~Kannathal, C.~M. Lim, and J.~S. Suri, ``Heart
  rate variability: a review,'' \emph{Medical and biological engineering and
  computing}, vol.~44, no.~12, pp. 1031--1051, 2006.

\bibitem{lecun1989backpropagation}
Y.~LeCun, B.~Boser, J.~S. Denker, D.~Henderson, R.~E. Howard, W.~Hubbard, and
  L.~D. Jackel, ``Backpropagation applied to handwritten zip code
  recognition,'' \emph{Neural computation}, vol.~1, no.~4, pp. 541--551, 1989.

\bibitem{vaswani2017attention}
A.~Vaswani, N.~Shazeer, N.~Parmar, J.~Uszkoreit, L.~Jones, A.~N. Gomez,
  {\L}.~Kaiser, and I.~Polosukhin, ``Attention is all you need,'' in
  \emph{Advances in neural information processing systems}, 2017, pp.
  5998--6008.

\bibitem{recommendation1988pulse}
I.~Switzerland, ``G. 711: Pulse code modulation (pcm) of voice frequencies,''
  \emph{ITU-T Recommendation G}, vol. 711, 1988.

\bibitem{test}
ANSI/AAMI, ``Testing and reporting performance results of cardiac rhythm and st
  segment measurement algorithms,'' \emph{American National Standards
  Institute, Inc. (ANSI), Association for the Advancement of Medical
  Instrumentation (AAMI), ANSI/AAMI/ISO EC57, 1998-(R)2008 (2008)}.

\bibitem{paszke2017automatic}
\BIBentryALTinterwordspacing
A.~Paszke, S.~Gross, S.~Chintala, G.~Chanan, E.~Yang, Z.~DeVito, Z.~Lin,
  A.~Desmaison, L.~Antiga, and A.~Lerer, ``Automatic differentiation in
  pytorch,'' in \emph{NIPS 2017 Workshop on Autodiff}, 2017. [Online].
  Available: \url{https://openreview.net/forum?id=BJJsrmfCZ}
\BIBentrySTDinterwordspacing

\bibitem{kingma2014adam}
D.~P. Kingma and J.~Ba, ``Adam: A method for stochastic optimization,''
  \emph{arXiv preprint arXiv:1412.6980}, 2014.

\bibitem{zhao2016emotion}
M.~Zhao, F.~Adib, and D.~Katabi, ``Emotion recognition using wireless
  signals,'' in \emph{Proceedings of the 22nd Annual International Conference
  on Mobile Computing and Networking}, 2016, pp. 95--108.

\bibitem{muller2007dynamic}
M.~M{\"u}ller, ``Dynamic time warping,'' \emph{Information retrieval for music
  and motion}, pp. 69--84, 2007.

\bibitem{likas2003global}
A.~Likas, N.~Vlassis, and J.~J. Verbeek, ``The global k-means clustering
  algorithm,'' \emph{Pattern recognition}, vol.~36, no.~2, pp. 451--461, 2003.

\bibitem{benini2014user}
A.~Benini, M.~Donati, F.~Iacopetti, and L.~Fanucci, ``User-friendly single-lead
  ecg device for home telemonitoring applications,'' in \emph{2014 8th
  International Symposium on Medical Information and Communication Technology
  (ISMICT)}.\hskip 1em plus 0.5em minus 0.4em\relax IEEE, 2014, pp. 1--5.

\bibitem{mahdy2018smart}
L.~N. Mahdy, K.~A. Ezzat, and Q.~Tan, ``Smart ecg holter monitoring system
  using smartphone,'' in \emph{2018 IEEE International Conference on Internet
  of Things and Intelligence System (IOTAIS)}.\hskip 1em plus 0.5em minus
  0.4em\relax IEEE, 2018, pp. 80--84.

\bibitem{yama2007development}
Y.~Yama, A.~Ueno, and Y.~Uchikawa, ``Development of a wireless capacitive
  sensor for ambulatory ecg monitoring over clothes,'' in \emph{2007 29th
  Annual International Conference of the IEEE Engineering in Medicine and
  Biology Society}.\hskip 1em plus 0.5em minus 0.4em\relax IEEE, 2007, pp.
  5727--5730.

\bibitem{petrenas2015modified}
A.~Petr{\.e}nas, V.~Marozas, G.~Jaru{\v{s}}evi{\v{c}}ius, and L.~S{\"o}rnmo,
  ``A modified lewis ecg lead system for ambulatory monitoring of atrial
  arrhythmias,'' \emph{Journal of electrocardiology}, vol.~48, no.~2, pp.
  157--163, 2015.

\bibitem{phan2015smartwatch}
D.~Phan, L.~Y. Siong, P.~N. Pathirana, and A.~Seneviratne, ``Smartwatch:
  Performance evaluation for long-term heart rate monitoring,'' in \emph{2015
  International symposium on bioelectronics and bioinformatics (ISBB)}.\hskip
  1em plus 0.5em minus 0.4em\relax IEEE, 2015, pp. 144--147.

\bibitem{aljuaid2020smartphone}
M.~Aljuaid, Q.~Marashly, J.~AlDanaf, I.~Tawhari, M.~Barakat, R.~Barakat,
  B.~Zobell, W.~Cho, M.~G. Chelu, and N.~F. Marrouche, ``Smartphone ecg
  monitoring system helps lower emergency room and clinic visits in
  post--atrial fibrillation ablation patients,'' \emph{Clinical Medicine
  Insights: Cardiology}, vol.~14, p. 1179546820901508, 2020.

\bibitem{fung2015electrocardiographic}
E.~Fung, M.-R. J{\"a}rvelin, R.~N. Doshi, J.~S. Shinbane, S.~K. Carlson, L.~P.
  Grazette, P.~M. Chang, R.~S. Sangha, H.~V. Huikuri, and N.~S. Peters,
  ``Electrocardiographic patch devices and contemporary wireless cardiac
  monitoring,'' \emph{Frontiers in physiology}, vol.~6, p. 149, 2015.

\bibitem{rachim2016wearable}
V.~P. Rachim and W.-Y. Chung, ``Wearable noncontact armband for mobile ecg
  monitoring system,'' \emph{IEEE transactions on biomedical circuits and
  systems}, vol.~10, no.~6, pp. 1112--1118, 2016.

\bibitem{ankhili2018washable}
A.~Ankhili, X.~Tao, C.~Cochrane, D.~Coulon, and V.~Koncar, ``Washable and
  reliable textile electrodes embedded into underwear fabric for
  electrocardiography (ecg) monitoring,'' \emph{Materials}, vol.~11, no.~2, p.
  256, 2018.

\bibitem{bouwstra2009smart}
S.~Bouwstra, W.~Chen, L.~Feijs, and S.~B. Oetomo, ``Smart jacket design for
  neonatal monitoring with wearable sensors,'' in \emph{2009 Sixth
  International Workshop on Wearable and Implantable Body Sensor
  Networks}.\hskip 1em plus 0.5em minus 0.4em\relax IEEE, 2009, pp. 162--167.

\bibitem{adib2015smart}
F.~Adib, H.~Mao, Z.~Kabelac, D.~Katabi, and R.~C. Miller, ``Smart homes that
  monitor breathing and heart rate,'' in \emph{Proceedings of the 33rd annual
  ACM conference on human factors in computing systems}, 2015, pp. 837--846.

\bibitem{zhang2020mtrack}
D.~Zhang, Y.~Hu, and Y.~Chen, ``Mtrack: Tracking multiperson moving
  trajectories and vital signs with radio signals,'' \emph{IEEE Internet of
  Things Journal}, vol.~8, no.~5, pp. 3904--3914, 2020.

\bibitem{dong2020cardiogram}
S.~Dong, Y.~Zhang, C.~Ma, Q.~Lv, C.~Li, and L.~Ran, ``Cardiogram detection with
  a millimeter-wave radar sensor,'' in \emph{2020 IEEE Radio and Wireless
  Symposium (RWS)}.\hskip 1em plus 0.5em minus 0.4em\relax IEEE, 2020, pp.
  127--129.

\bibitem{ha2020contactless}
U.~Ha, S.~Assana, and F.~Adib, ``Contactless seismocardiography via deep
  learning radars,'' in \emph{Proceedings of the 26th Annual International
  Conference on Mobile Computing and Networking}, 2020, pp. 1--14.

\bibitem{quinn2014cardiac}
T.~A. Quinn, P.~Kohl, and U.~Ravens, ``Cardiac mechano-electric coupling
  research: fifty years of progress and scientific innovation,'' \emph{Progress
  in biophysics and molecular biology}, vol. 115, no. 2-3, pp. 71--75, 2014.

\bibitem{kohl2011cardiac}
P.~Kohl, F.~Sachs, and M.~R. Franz, \emph{Cardiac mechano-electric coupling and
  arrhythmias}.\hskip 1em plus 0.5em minus 0.4em\relax Oxford University Press,
  2011.

\bibitem{chen2016time}
C.~Chen, Y.~Han, Y.~Chen, F.~Zhang, and K.~R. Liu, ``Time-reversal indoor
  positioning with centimeter accuracy using multi-antenna wifi,'' in
  \emph{2016 IEEE Global Conference on Signal and Information Processing
  (GlobalSIP)}.\hskip 1em plus 0.5em minus 0.4em\relax IEEE, 2016, pp.
  1022--1026.

\bibitem{chen2016achieving}
C.~Chen, Y.~Chen, Y.~Han, H.-Q. Lai, and K.~R. Liu, ``Achieving
  centimeter-accuracy indoor localization on wifi platforms: A frequency
  hopping approach,'' \emph{IEEE Internet of Things Journal}, vol.~4, no.~1,
  pp. 111--121, 2016.

\bibitem{chen2020speednet}
Y.~Chen, H.~Deng, D.~Zhang, and Y.~Hu, ``Speednet: Indoor speed estimation with
  radio signals,'' \emph{IEEE Internet of Things Journal}, vol.~8, no.~4, pp.
  2762--2774, 2020.

\bibitem{hsu2017extracting}
C.-Y. Hsu, Y.~Liu, Z.~Kabelac, R.~Hristov, D.~Katabi, and C.~Liu, ``Extracting
  gait velocity and stride length from surrounding radio signals,'' in
  \emph{Proceedings of the 2017 CHI Conference on Human Factors in Computing
  Systems}, 2017, pp. 2116--2126.

\bibitem{li2021towards}
Y.~Li, D.~Zhang, J.~Chen, J.~Wan, D.~Zhang, Y.~Hu, Q.~Sun, and Y.~Chen,
  ``Towards domain-independent and real-time gesture recognition using mmwave
  signal,'' \emph{to appear in IEEE Transactions on Mobile Computing}, 2022,
  DOI: 10.1109/TMC.2022.3207570.

\bibitem{zhang2021unsupervised}
B.-B. Zhang, D.~Zhang, Y.~Li, Y.~Hu, and Y.~Chen, ``Unsupervised domain
  adaptation for device-free gesture recognition,'' \emph{arXiv preprint
  arXiv:2111.10602}, 2021.

\bibitem{zhao2018rf}
M.~Zhao, Y.~Tian, H.~Zhao, M.~A. Alsheikh, T.~Li, R.~Hristov, Z.~Kabelac,
  D.~Katabi, and A.~Torralba, ``Rf-based 3d skeletons,'' in \emph{Proceedings
  of the 2018 Conference of the ACM Special Interest Group on Data
  Communication}, 2018, pp. 267--281.

\bibitem{zhao2018through}
M.~Zhao, T.~Li, M.~Abu~Alsheikh, Y.~Tian, H.~Zhao, A.~Torralba, and D.~Katabi,
  ``Through-wall human pose estimation using radio signals,'' in
  \emph{Proceedings of the IEEE Conference on Computer Vision and Pattern
  Recognition}, 2018, pp. 7356--7365.

\bibitem{quinn2021cardiac}
T.~A. Quinn and P.~Kohl, ``Cardiac mechano-electric coupling: acute effects of
  mechanical stimulation on heart rate and rhythm,'' \emph{Physiological
  reviews}, vol. 101, no.~1, pp. 37--92, 2021.

\bibitem{he2020wifi}
Y.~He, Y.~Chen, Y.~Hu, and B.~Zeng, ``Wifi vision: sensing, recognition, and
  detection with commodity mimo-ofdm wifi,'' \emph{IEEE Internet of Things
  Journal}, vol.~7, no.~9, pp. 8296--8317, 2020.

\bibitem{xu2017trieds}
{Xu, Qinyi and Chen, Yan and Wang, Beibei and Liu, KJ Ray}, ``Trieds: Wireless
  events detection through the wall,'' \emph{IEEE Internet of Things Journal},
  vol.~4, no.~3, pp. 723--735, 2017.

\bibitem{xu2017radio}
Q.~Xu, Y.~Chen, B.~Wang, and K.~R. Liu, ``Radio biometrics: Human recognition
  through a wall,'' \emph{IEEE Internet of Things Journal}, vol.~12, no.~5, pp.
  1141--1155, 2017.

\bibitem{han2016enabling}
Y.~Han, Y.~Chen, B.~Wang, and K.~R. Liu, ``Enabling heterogeneous connectivity
  in internet of things: A time-reversal approach,'' \emph{IEEE Internet of
  Things Journal}, vol.~3, no.~6, pp. 1036--1047, 2016.

\bibitem{zhang2019calibrating}
D.~Zhang, Y.~Hu, Y.~Chen, and B.~Zeng, ``Calibrating phase offsets for
  commodity wifi,'' \emph{IEEE Systems Journal}, vol.~14, no.~1, pp. 661--664,
  2019.

\bibitem{zhang2018multitarget}
D.~Zhang, Y.~He, X.~Gong, Y.~Hu, Y.~Chen, and B.~Zeng, ``Multitarget aoa
  estimation using wideband lfmcw signal and two receiver antennas,''
  \emph{IEEE Transactions on Vehicular Technology}, vol.~67, no.~8, pp.
  7101--7112, 2018.

\bibitem{chen2019residual}
Y.~Chen, X.~Su, Y.~Hu, and B.~Zeng, ``Residual carrier frequency offset
  estimation and compensation for commodity wifi,'' \emph{IEEE Transactions on
  Mobile Computing}, vol.~19, no.~12, pp. 2891--2902, 2019.

\bibitem{holoborodko2014noise}
\BIBentryALTinterwordspacing
``Noise robust differentiators for second derivative estimation,'' accessed
  August 4, 2022. [Online]. Available:
  \url{http://www.holoborodko.com/pavel/downloads/NoiseRobustSecondDerivative}
\BIBentrySTDinterwordspacing

\bibitem{bers2002cardiac}
D.~M. Bers, ``Cardiac excitation--contraction coupling,'' \emph{Nature}, vol.
  415, no. 6868, pp. 198--205, 2002.

\bibitem{bers2001excitation}
D.~Bers, \emph{Excitation-contraction coupling and cardiac contractile
  force}.\hskip 1em plus 0.5em minus 0.4em\relax Springer Science \& Business
  Media, 2001, vol. 237.

\bibitem{gurev2011models}
V.~Gurev, T.~Lee, J.~Constantino, H.~Arevalo, and N.~A. Trayanova, ``Models of
  cardiac electromechanics based on individual hearts imaging data,''
  \emph{Biomechanics and modeling in mechanobiology}, vol.~10, no.~3, pp.
  295--306, 2011.

\bibitem{christoph2020inverse}
J.~Christoph and J.~Lebert, ``Inverse mechano-electrical reconstruction of
  cardiac excitation wave patterns from mechanical deformation using deep
  learning,'' \emph{Chaos: An Interdisciplinary Journal of Nonlinear Science},
  vol.~30, no.~12, p. 123134, 2020.

\bibitem{christoph2018electromechanical}
J.~Christoph, M.~Chebbok, C.~Richter, J.~Schr{\"o}der-Schetelig, P.~Bittihn,
  S.~Stein, I.~Uzelac, F.~H. Fenton, G.~Hasenfu{\ss}, R.~Gilmour~Jr
  \emph{et~al.}, ``Electromechanical vortex filaments during cardiac
  fibrillation,'' \emph{Nature}, vol. 555, no. 7698, pp. 667--672, 2018.

\bibitem{provost2011electromechanical}
J.~Provost, V.~T.-H. Nguyen, D.~Legrand, S.~Okrasinski, A.~Costet, A.~Gambhir,
  H.~Garan, and E.~E. Konofagou, ``Electromechanical wave imaging for
  arrhythmias,'' \emph{Physics in Medicine \& Biology}, vol.~56, no.~22, p.~L1,
  2011.

\bibitem{gulrajani1998forward}
R.~M. Gulrajani, ``The forward and inverse problems of electrocardiography,''
  \emph{IEEE Engineering in Medicine and Biology Magazine}, vol.~17, no.~5, pp.
  84--101, 1998.

\bibitem{boulakia2010mathematical}
M.~Boulakia, S.~Cazeau, M.~A. Fern{\'a}ndez, J.-F. Gerbeau, and N.~Zemzemi,
  ``Mathematical modeling of electrocardiograms: a numerical study,''
  \emph{Annals of biomedical engineering}, vol.~38, no.~3, pp. 1071--1097,
  2010.

\bibitem{lines2003mathematical}
G.~Lines, M.~Buist, P.~Grottum, A.~Pullan, J.~Sundnes, and A.~Tveito,
  ``Mathematical models and numerical methods for the forward problem in
  cardiac electrophysiology,'' \emph{Computing and Visualization in Science},
  vol.~5, no.~4, pp. 215--239, 2003.

\bibitem{gurev2012mechanisms}
V.~Gurev, K.~Tavakolian, J.~Constantino, B.~Kaminska, A.~P. Blaber, and N.~A.
  Trayanova, ``Mechanisms underlying isovolumic contraction and ejection peaks
  in seismocardiogram morphology,'' \emph{Journal of medical and biological
  engineering}, vol.~32, no.~2, p. 103, 2012.

\bibitem{laurin20153d}
A.~Laurin, S.~Imperiale, P.~Moireau, A.~Blaber, and D.~Chapelle, ``A 3d model
  of the thorax for seismocardiography,'' in \emph{2015 Computing in Cardiology
  Conference (CinC)}.\hskip 1em plus 0.5em minus 0.4em\relax IEEE, 2015, pp.
  465--468.

\bibitem{tiref}
``Texas instruments awr1843boost,''
  \url{https://www.ti.com/lit/ug/spruim4b/spruim4b.pdf?ts=1660539830441},
  accessed August 4, 2022.

\end{thebibliography}
\bibliographystyle{unsrt}

\end{document}